\documentclass[twocolumn,epjc3]{svjour3mod}  
\journalname{Eur. Phys. J. C}
\usepackage[utf8]{inputenc}
\usepackage{epsfig,amsfonts,amssymb,latexsym,amsmath,bm,color,array,epsfig,graphics,dsfont}
\usepackage{enumitem}
\usepackage{amsmath}
\usepackage{amssymb}
\usepackage{physics}
\usepackage{lastpage}
\usepackage{graphicx}
\usepackage{subfig}
\usepackage{hyperref}
\usepackage{balance,cite}
\usepackage{booktabs}
\hypersetup{colorlinks=true,linkcolor=blue,citecolor=blue,menucolor=black,urlcolor=blue,filecolor=blue}
\newcommand{\kpi}{\pi K}
\newcommand{\keta}{\eta K}
\newcommand{\ketap}{\eta^\prime K}
\newcommand{\VR}{V_\text{R}}
\newcommand{\TR}{T_\text{R}}
\newcommand{\tR}{t_\text{R}}
\newcommand{\disc}{\text{disc}\,}
\newcommand{\FF}{f_\text{s}}
\newcommand{\Ka}{K_0^*(700)}
\newcommand{\Kb}{K_0^*(1430)}
\newcommand{\Kc}{K_0^*(1950)}
\newcommand{\Kd}{K^*(892)}
\newcommand{\Ke}{K^*(1410)}
\newcommand{\sm}{s_\text{m}}
\newcommand{\mpi}{M_\pi}
\newcommand{\mk}{M_K}
\newcommand{\meta}{M_\eta}
\newcommand{\metap}{M_{\eta^\prime}}
\newcommand{\mtau}{m_\tau}
\newcommand{\mr}[1]{\widetilde{M}_{#1}}
\newcommand{\gammar}[1]{\widetilde{\Gamma}_{#1}}
\newcommand{\pk}{p_K}
\newcommand{\ppi}{p_\pi}
\newcommand{\beq}{\begin{equation}}
\newcommand{\eeq}{\end{equation}}
\newcommand{\K}{K}
\newcommand{\Id}{\mathds{1}}

\newcommand{\GeV}{\,\text{GeV}}
\newcommand{\MeV}{\,\text{MeV}}

\allowdisplaybreaks[1]

\begin{document}
\emergencystretch 3em
\title{On the scalar $\boldsymbol{\pi K}$ form factor beyond the elastic region}
\author{ 
L. von Detten\thanksref{add1} \and 
F.  No\"{e}l\thanksref{add1,add2} \and 
C. Hanhart\thanksref{add1} \and
M. Hoferichter\thanksref{add2} \and
B. Kubis\thanksref{add3} 
}       
\institute{Forschungszentrum J\"ulich, Institute for Advanced Simulation, Institut f\"ur Kernphysik, and
J\"ulich Center for Hadron Physics, 52425 J\"ulich, Germany \label{add1}
\and
Albert Einstein Center for Fundamental Physics, Institute for Theoretical Physics, University of Bern, Sidlerstrasse 5, 3012 Bern, Switzerland \label{add2}
\and
Helmholtz-Institut f\"ur Strahlen- und Kernphysik and Bethe Center for Theoretical Physics, Universit\"at Bonn, 53115 Bonn, Germany \label{add3}
}
\date{}
\maketitle
\begin{abstract}
Pion--kaon ($\pi K$) pairs occur frequently as final states in heavy-particle decays. 
A consistent treatment of $\pi K$ scattering and production amplitudes over a wide energy range is therefore mandatory for multiple applications:
in Standard Model tests; to describe crossed channels in the quest for exotic hadronic states; and for an improved spectroscopy of excited kaon resonances.
In the elastic region, the phase shifts of $\pi K$ scattering in a given partial wave are related to the phases of the respective $\pi K$ form factors by Watson's theorem. 
Going beyond that, we here construct a representation of the scalar $\pi K$  form factor that includes inelastic effects via resonance exchange, while fulfilling all constraints from $\pi K$ scattering and maintaining the correct analytic structure. 
As a first application, we consider the decay ${\tau\to K_S\pi\nu_\tau}$, in particular, we study to which extent the $S$-wave $\Kb$ and the $P$-wave $\Ke$ resonances can be differentiated and provide an improved estimate of the $CP$ asymmetry produced by a tensor operator. 
Finally, we extract the pole parameters of the $\Kb$ and $\Kc$ resonances via Pad\'e approximants, $\sqrt{s_{\Kb}}=[1408(48)-i\, 180(48)]\MeV$ and $\sqrt{s_{\Kc}}=[1863(12)-i\,136(20)]\MeV$, as well as the pole residues.  A generalization of the method also allows us to formally define a branching fraction for ${\tau\to \Kb \nu_\tau}$ in terms of the corresponding residue, leading to the upper limit 
    ${\text{BR}(\tau\to \Kb \nu_\tau)<1.6 \times 10^{-4}}$.
\end{abstract}
\section{Introduction}

At low energies, the $\pi K$ $S$-wave of isospin $1/2$ is characterized by the interplay of low-energy theorems induced by the chiral structure of QCD~\cite{Bernard:1990kw,Bijnens:2004bu} and a relatively close-by pole located deep in the complex plane called the $\kappa$ or $K^*(700)$~\cite{Buettiker:2003pp,DescotesGenon:2006uk,Pelaez:2016klv,Pelaez:2020uiw,Pelaez:2020gnd,Pelaez:2021dak}. The properties of the $\kappa$ cannot be described by a simple Breit--Wigner (BW) model, but require the proper consideration of the analytic structure, most conveniently implemented in the framework of dispersion relations. Given that the $\pi K$ $S$-wave effectively stays elastic well beyond $1\GeV$, with the first excited resonance, the $\Kb$, still predominantly coupling to the $\kpi$ channel, the $\kappa$ properties are thus largely encoded in the $S$-wave phase shift, although the full dispersive analysis involves other partial waves as well as the crossed reaction ${\pi\pi\to\bar K K}$~\cite{Buettiker:2003pp,Pelaez:2018qny,Pelaez:2020gnd}.
While $\kpi$ scattering thus serves as the simplest probe of the strangeness sector of the QCD spectrum, its impact extends far beyond, with more complicated processes such as ${\gamma K\to\pi K}$~\cite{Dax:2020dzg}, $K_{\ell 4}$ decays~\cite{Colangelo:2015kha}, $D$-meson decays such as $D\to\pi\pi K$~\cite{Niecknig:2015ija,Niecknig:2017ylb}, or even reactions involving nucleons~\cite{Ditsche:2012fv,Hoferichter:2015hva} depending on $\pi K$ amplitudes as input. 

Moreover, the same principles of unitarity and analyticity upon which modern analyses of $\pi K$ scattering are based imply a relation to the corresponding form factors. In the crossed reaction ${\pi\pi\to\bar K K}$ this connection determines scalar meson~\cite{Donoghue:1990xh} and nucleon~\cite{Hoferichter:2012wf} form factors via a coupled-channel $T$-matrix, while the $\kpi$ form factors of a given partial wave are directly related to the respective $\kpi$ scattering amplitudes via Watson's theorem~\cite{Watson:1954uc}, which states that the phases coincide in the elastic region.  
The $S$- and $P$-wave $\kpi$ form factors are relevant for analyses of $K_{\ell 3}$~\cite{Bernard:2006gy,Bernard:2009zm,Abouzaid:2009ry} and ${\tau\to K_S\pi\nu_\tau}$ decays~\cite{Moussallam:2007qc,Boito:2008fq,Boito:2010me,Bernard:2011ae,Antonelli:2013usa}, where the $\tau$ spectrum probes the region of parameter space in which an elastic approximation no longer applies. Extensions of the simple Omn\`es representation~\cite{Omnes:1958hv} are thus required. For the $P$-wave, inelastic effects are typically included in resonance chiral theory (RChT)~\cite{Ecker:1988te} via the $\Ke$, providing an extended parameterization of the phase shift to be used in the Omn\`es factor or by feeding the corresponding amplitudes into a unitarization scheme such as the $N/D$ method~\cite{Oller:1998zr,Jamin:2000wn}.
The latter is hard to handle, however, since it is difficult to prevent its high-order polynomials from generating unphysical poles~\cite{Noel:2020vpo}.
Moreover, for the $S$-wave, the effect of inelasticities in  ${\tau\to K_S\pi\nu_\tau}$ is usually neglected apart from a generous variation of the unknown phase of the form factor, leading to an Omn\`es representation that, besides constraints from the Callan--Treiman low-energy theorem~\cite{Callan:1966hu,Dashen:1969bh,Gasser:1984ux,Bijnens:2007xa,Kastner:2008ch}, essentially involves a subtracted version of the elastic solution. 

Extending the applicability range of form factor parameterizations by an improved treatment of inelastic effects has become increasingly pressing in recent years. First, the size of the $CP$ asymmetry in ${\tau\to K_S\pi\nu_\tau}$ generated by a tensor operator was shown to be solely determined by inelastic effects~\cite{Cirigliano:2017tqn}, due to a cancellation of the elastic contribution that follows from Watson's theorem. In addition, control over inelastic effects would be required to describe ${D\to \pi K\ell\nu_\ell}$~\cite{Ablikim:2015mjo} and future measurements of ${B\to \pi K\ell\nu_\ell}$ or ${B\to \pi K \ell\ell}$, or as subamplitudes 
for the calculation of heavy-meson Dalitz plots, which are often described in a simplified manner in terms of $\kpi$ form factors~\cite{Oller:2004xm,ElBennich:2006yi,ElBennich:2009da,Boito:2009qd,Boito:2017jav}. For the latter application, the amplitudes are described by the same form factors if the impact of hadronic spectator particles is neglected, and in this case variants of the scalar form factor have been constructed that include inelastic effects by a coupled-channel treatment of $\pi K$ and $\eta' K$~\cite{ElBennich:2009da}.

Also in the hunt for exotic hadrons, controlled $\pi K$ amplitudes are very valuable. For example, at Belle and LHCb the $Z_c(4430)$ was discovered in the reaction ${B\to \psi'\pi K}$ in the $\psi'\pi$ subsystem~\cite{Mizuk:2009da,Aaij:2014jqa}. The signal became visible through the observation that the $\pi K$ amplitudes in the crossed channel were not able to describe the $\psi'\pi$ distribution. Since in such crossed amplitudes the individual partial waves interfere with each other, a high control especially of their phases is mandatory. 
Finally, to get access to the spectrum of kaon resonances and in particular their pole parameters, employing amplitudes consistent with analyticity and unitarity is necessary.

In this paper, we propose a parameterization for the $S$-wave $\pi K$ form factor that has the
proper low-energy behavior and at the same time allows for an inclusion of resonances and inelasticities at higher energies. We follow the strategy from Ref.~\cite{Ropertz:2018stk} (originally proposed in Ref.~\cite{Hanhart:2012wi} for the pion vector form factor), describing inelastic effects via resonances akin to the isobar model, but in such a way that at low energies the elastic Omn\`es parameterization is reproduced and the correct analytic structure remains preserved. Accordingly, we assume that the inelastic contributions can be understood as proceeding via resonances, as supported by the phenomenological success of the isobar model.  
The analogous representation derived in Ref.~\cite{Ropertz:2018stk} then allowed for an analysis of the complete kinematic range of the ${B_{s}\to J/\psi \pi \pi}$ and ${B_{s}\to J/\psi \bar KK}$ spectra, extending the previous high-quality description in a restricted range of ${\pi\pi}$ and ${\bar KK}$ invariant masses~\cite{Daub:2015xja}. In particular, the properties of the higher $S$-wave resonances could be extracted. 

In this work, we first establish a similar formalism for the $\pi K$ system, see Sect.~\ref{sec:formalism}, with the input from $\pi K$ scattering data discussed in Sects.~\ref{sec:scattering_data} and~\ref{sec:fit_scattering}. As applications, we consider the ${\tau\to K_S\pi\nu_\tau}$ spectrum in Sect.~\ref{sec:tau}, including an improved prediction for the $CP$ asymmetry produced by a tensor operator, and extract the resonance parameters of $\Kb$ and $\Kc$ in Sect.~\ref{sec:poles}, where the residue describing the coupling to the weak current allows us to formally define the branching fraction for ${\tau\to \Kb \nu_\tau}$.   
Our conclusions are given in Sect.~\ref{sec:conclusions}. 

\section{Formalism}\label{sec:formalism}

As mentioned above, we aim at a parameterization of the $\kpi$ isospin-$1/2$  $S$-wave scattering amplitude that at low energies matches smoothly onto elastic $\kpi$ scattering given by the input phase shift $\delta_0$, and at the same time allows for the inclusion of resonances and inelastic channels, most importantly the ${\ketap}$ channel, at higher energies---the $\eta K$ channel turns out to largely decouple. 
Thus, in the energy range we study, two channels are sufficient and we therefore present the formalism in a two-channel formulation, although an extension to more channels is straightforward. 
To derive an expression for the $T$-matrix that fulfills the mentioned criteria, we start from the Bethe--Salpeter equation, which in matrix form in channel space reads
\begin{equation}
	T_{if}=V_{if}+V_{im} G_{mm} T_{mf}\,,
	\label{eq:T_bethe_salpeter}
\end{equation}
where ${V_{if} \in \mathbb{R}}$ denotes the interaction potential between the initial channel $i$ and final channel $f$ and $G_{mm}$ denotes the loop operator, which provides the free propagation of the intermediate particles of channel $m$.
For two-particle states, its discontinuity is given by ${\disc G_{mm}=2i \rho_m}$, where $\rho_m$ denotes the two-body phase space in channel $m$,
\begin{equation}
    \rho_m(s)=\frac{\lambda^{\frac{1}{2}}\qty(s,\big(m^{(m)}_i\big)^2,\big(m^{(m)}_j\big)^2)}{16 \pi s}\,,
\end{equation} 
where $\lambda$ is the K\"all\'en function 
\begin{equation}
\lambda(a,b,c)=a^2+b^2+c^2-2(a b+a c+b c)\,.
\end{equation} 
To proceed  we follow the general concepts of the so-called two-potential formalism~\cite{Nakano:1982bc}, which calls for splitting the scattering  potential $V$ into two pieces,
\begin{equation}
V= V_0 + V_\text{R}\,.
\label{eq:V_split}
\end{equation}
This allows for a corresponding splitting of the $T$-matrix
\begin{equation}
T=T_0+\TR\,,
\label{eq:T_split}
\end{equation}
where $T_0$ fulfills the Bethe--Salpeter equation
that has $V_0$ as input, ${T_0=V_0+V_0 G T_0}$.
As will be demonstrated below, the explicit form of $V_0$ is never needed: all quantities necessary to express the full scattering $T$-matrix  and the scalar form factor can be calculated from the scattering phase shift $\delta_0$ directly. 
$T_0$ fixes the low-energy behavior of the model, while $\TR$ incorporates the high-energy resonant behavior via $V_\text{R}$. 
In the case of $\kpi$ scattering studied here, we assume $T_0$ to be purely elastic. The additional channel couples through the resonance exchange in $T_\text{R}$ only. We may therefore write
\begin{equation}
T_0=\begin{pmatrix}
\tfrac{1}{\rho_1}\sin{\delta_0} e^{i\delta_0} & 0\\
0 & 0
\end{pmatrix}\,.
\end{equation}
Clearly, the assumption that all higher channels couple via resonances introduces some model dependence, which, however, is backed by phenomenology~\cite{Anisovich:2002ij,Klempt:2007cp,Battaglieri:2014gca}. 
We furthermore define the vertex function ${\Omega=\Id+T_0 G}$. Its discontinuity is given by 
\begin{equation}
{\disc \Omega_{if} =2 i  (T_0^*)_{im} \rho_m \Omega_{mf}}\,,
\end{equation}
 which matches that of an Omn\`es function~\cite{Omnes:1958hv} calculated from $T_0$. Thus we can express $\Omega$ via a dispersion integral over the input phase $\delta_0$, 
\begin{equation}
\label{eq:Omnes}
\Omega=\begin{pmatrix}
\Omega_{11} & 0\\
0 & 1
\end{pmatrix}\,, \; \Omega_{11}=\exp \qty(\frac{s}{\pi} \int_{s_\text{th}}^\infty \text{d}z \frac{\delta_0(z)}{z(z-s)})\,.
\end{equation}
Note that in order to render the integral well defined, the phase $\delta_0$ needs to be continued up to infinite energies.
How this is done in practice is discussed below.
Plugging Eqs.~\eqref{eq:V_split} and~\eqref{eq:T_split} into Eq.~\eqref{eq:T_bethe_salpeter}, one
finds after some algebra the defining equation for $\tR$, 
\begin{equation}
\tR=V_\text{R}+V_\text{R} \Sigma \tR \,, 
\label{eq:tr_bethe_salpeter}
\end{equation}
which is related to $\TR$ via ${\TR=\Omega \tR \Omega^T}$.
The so-called dressed loop operator or self energy ${\Sigma=G\Omega}$ incorporates the effects contained in $T_0$ into the propagation of the two-meson states as demanded by unitarity. 
It can be expressed as a once-subtracted dispersion integral
\begin{equation}
\Sigma_{ij}(s)=\frac{s}{2 \pi i } \int_{s_\text{th}}^{\infty} \text{d}z \frac{\text{disc}\Sigma_{ij}(z)}{z(z-s)}\,,
\label{eq:selfenergy}
\end{equation}
with its discontinuity given by 
\begin{equation}
{\disc \Sigma_{if}=\Omega^\dagger_{im}  \disc G_{mm} \Omega_{mf}}\,.
\end{equation}
The subtraction constant is reabsorbed into the potential $V_\text{R}$. Such manipulations are justified as the formalism has not made any assumptions about the form of $V_\text{R}$ besides it being real and having poles at the bare resonance masses $\mr{(r)}$.
The simplest parameterization of this kind is
\begin{equation}
\overline{V}_\text{R}(s)_{ij}=-\sum_{r}\frac{g_i^{(r)} g_j^{(r)}}{s-\mr{(r)}^2} \,,
\end{equation}
where the $g_i^{(r)}$ denote the bare couplings of the resonance $r$ to channel $i$. 
The bare parameters introduced here should not be confused with the physical parameters introduced in Sect.~\ref{sec:poles}. 
To reduce the impact of $\VR$ at lower energies, the potential is subtracted at some properly chosen point $s_0$, resulting in
\begin{align}\nonumber
V_\text{R}(s)_{ij}&=\overline{V}_\text{R}(s)_{ij}-\overline{V}_\text{R}(s_0)_{ij}\\
&= \sum_r g_i^{(r)}\frac{s-s_0}{\qty(s-\mr{(r)}^2)\qty(s_0-\mr{(r)}^2)} g_j^{(r)}\,. \label{eq:VR}
\end{align}
Solving Eq.~\eqref{eq:tr_bethe_salpeter} for $\tR$, the full scattering $T$-matrix  is given by
\begin{equation}
T = T_0 + T_\text{R} = T_0 + \Omega \left[ \Id-V_\text{R} \Sigma \right]^{-1} V_\text{R} \Omega^\text{T}\,,
\label{eq:T_theory}
\end{equation}
with $V_\text{R}$ as defined in Eq.~\eqref{eq:VR}.

We can further parameterize the $\kpi$ production mechanism by adapting the $P$-vector formalism of Ref.~\cite{Aitchison:1972ay} (see also the resonance review of Ref.~\cite{Zyla:2020zbs}). 
The scalar form factor $\FF(s)$ is then expressed as
\begin{equation}
(\FF)_i=M_i + T_{im} G_{mm} M_m \,,
\label{eq:FF_general}
\end{equation}
where $M$ is some properly chosen source term. 
Under the assumption that $M$ does not contain any left-hand cuts, plugging Eq.~\eqref{eq:T_theory} into Eq.~\eqref{eq:FF_general} yields
\begin{equation}
\FF(s) = \Omega(s) \left[  \Id-V_\text{R}(s) \Sigma(s)  \right]^{-1} M(s) \,,
\label{eq:FF}
\end{equation} 
where $M$ is now a reparameterized source term, which can be written as
\begin{align}\nonumber
M_i &= \sum_{k=0}^{k_{\rm max}} c_i^{(k)}s^k \\
& \qquad - \sum_{r} g_i^{(r)} \frac{s-s_0}{\qty(s-\mr{(r)}^2)\qty(s_0-\mr{(r)}^2)} \alpha^{(r)} \,.
\label{eq:FF_source}
\end{align}
The coefficients $c_i^{(k)}$ and the resonance couplings $\alpha^{(r)}$ depend on the source.
A method to generalize the formalism to also allow for left-hand cuts is provided by the Khuri--Treiman formalism~\cite{Khuri:1960zz}. For a recent calculation of this kind where the amplitudes of Ref.~\cite{Ropertz:2018stk} were employed, see Ref.~\cite{Baru:2020ywb}.

\section{Scattering data and input phase}
\label{sec:scattering_data}

Most of the data on $\kpi$ scattering were obtained in the 1970s and 1980s.\footnote{New $\kpi$ scattering data are planned to be taken by a neutral-kaon-beam experiment at Jefferson Lab~\cite{Amaryan:2020xhw}.} Various experiments~\cite{Bakker:1970wg,Cho:1970fb,Estabrooks:1977xe,Jongejans:1973pn,Linglin:1973ci} obtained data for the phase shift of the isospin-${3}/{2}$ wave in the elastic regime from kaon--nucleon reactions using protons, neutrons, and deuterons with $K^\pm \pi^\pm$ in the final states.
The isospin-${1}/{2}$ wave, however, can only be measured in combination with the isospin-${3}/{2}$ wave, so that we mainly focus on the combination of both, which in terms of the $T$-matrices is expressed by
\begin{equation}
\hat{T}_{if}= \rho_i \qty(T^{{\frac{1}{2}}}+T^{{\frac{3}{2}}}/2)_{if}\,. \label{eq:That}
\end{equation}
Studies of the reaction ${K^- {p} \rightarrow K^-\pi^+ {n}}$ performed in Ref.~\cite{Aston:1987ir} resulted in data for argument and modulus of this isospin combination in the $\kpi$ channel up to about $2.5\GeV$, which we use to fix the free parameters of the resonance potential. 

\begin{figure}[tp]
	\includegraphics[width=\linewidth,trim = 30pt 30pt 0pt 0pt]{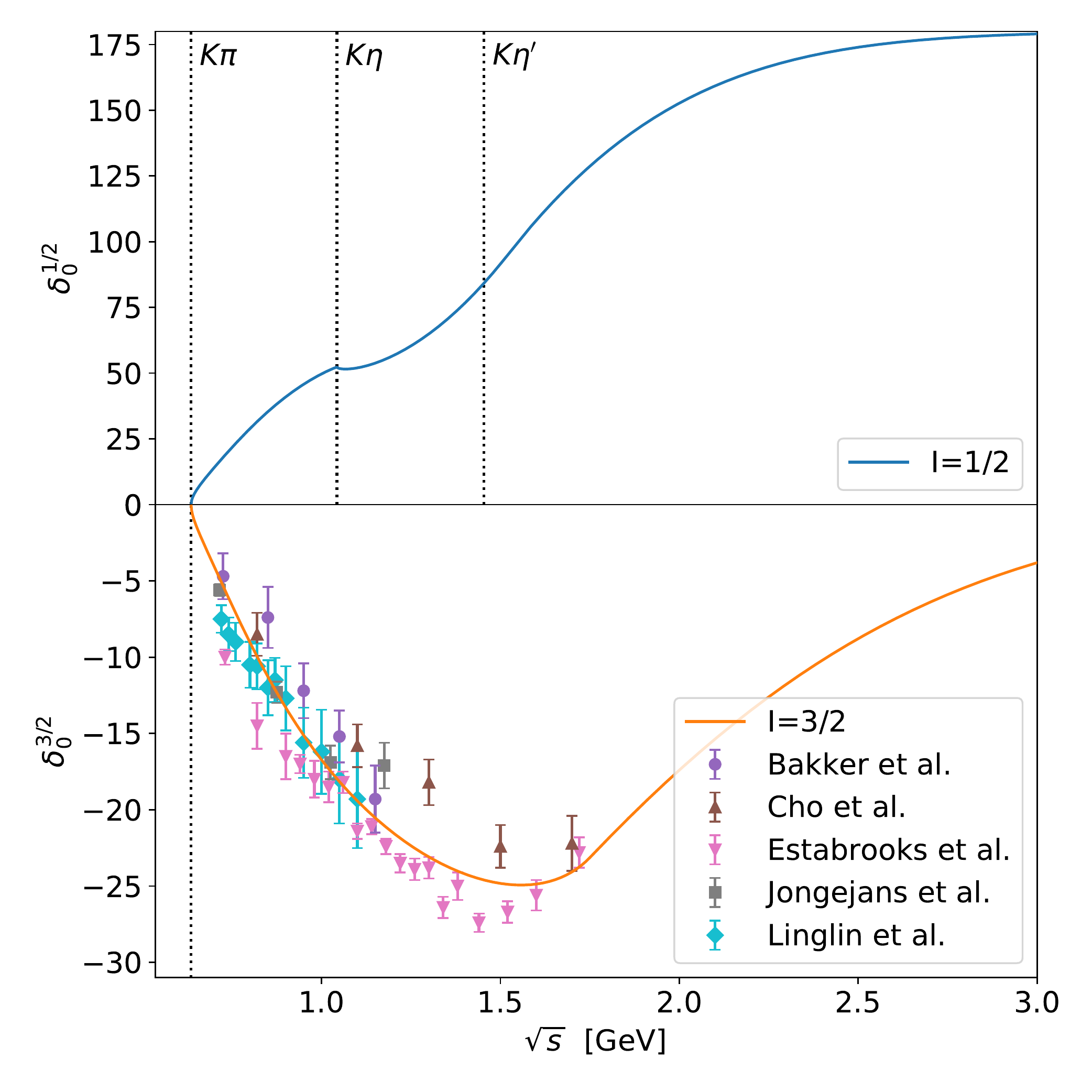}
	\centering
	\caption{Isospin-${1}/{2}$ and -${3}/{2}$ phase shifts including their high-energy extension. The latter is compared to the data from Bakker et al.~\cite{Bakker:1970wg}, Cho et al.~\cite{Cho:1970fb}, Estabrooks et al.~\cite{Estabrooks:1977xe}, Jongejans et al.~\cite{Jongejans:1973pn}, and Linglin et al.~\cite{Linglin:1973ci}.}
	\label{fig:Input_phase}
\end{figure}

For the low-energy phase shift $\delta_0$ we use the results obtained in Ref.~\cite{Pelaez:2016tgi}. 
Using forward dispersion relations to constrain the parameters, in that  
work the authors found a parameterization of the isospin-${1}/{2}$ and  -${3}/{2}$ waves 
up to $1.6\GeV$  and $1.8\GeV$, respectively. In the elastic regime it is based on a conformal expansion of the phase shifts, while inelastic background and resonance contributions are modeled by products of functions consistent with unitarity. 
For the input phase $\delta_0$ we reduce the parameterization to be purely elastic.
In addition, we remove the resonance contributions from the parameterization of the phase above the $\keta$ threshold, as higher resonances will be included via the resonance potential $V_\text{R}$. Thus, we use the parameterization provided in Ref.~\cite{Pelaez:2016tgi} as the input phase $\delta_0$ below the $\keta$ threshold, and set the parameters $G_1$ and $G_2$ to zero in the resonance terms $S_r^1$ and $S_r^2$ of Eq.~(16) in Ref.~\cite{Pelaez:2016tgi} above. This procedure makes a small cusp at the $\keta$ threshold more visible 
(cf.\ Fig.~\ref{fig:Input_phase}).
Since $T_0$ needs to be known in the full energy range and $\delta_0$ formally even up to infinite energies to allow one to evaluate the Omn\`es integral of Eq.~\eqref{eq:Omnes}, 
the phases needs to be continued smoothly up to high energies.
We force them to approach integer multiples of $\pi$, employing
\begin{align}
\delta_0(s)&=L-\big(L- \delta_0(\sm)\big) \exp\left( -\frac{(s{-}\sm) \delta_0^\prime(\sm)}{L{-}\delta_0(\sm)} \right)
\end{align}
for ${s>\sm}$. 
Here $L$ denotes the asymptotic limit of the phase shift  
and $\delta_0^\prime(s)$ its derivative $\text{d}{\delta_0(s)}/{\text{d}s}$.
As the isospin-${3}/{2}$ wave is purely elastic over a wide energy range and contains no resonances, which would be exotic due to their quantum numbers, the phase shift can simply be guided towards 0 as ${\TR^{\frac{3}{2}}=0}$. 

As shown in Fig.~\ref{fig:Input_phase}, the available data for the isospin-${3}/{2}$ wave of Refs.~\cite{Estabrooks:1977xe,Bakker:1970wg,Cho:1970fb,Linglin:1973ci,Jongejans:1973pn} are not mutually consistent, but the parameterization of Ref.~\cite{Pelaez:2016tgi} describes them quite decently. As data are only available up to $1.72\GeV$, we choose ${\sqrt{\sm}=1.75\GeV}$ as a matching point for the isospin-${3}/{2}$ wave. 
The isospin-${1}/{2}$ wave on the other hand is guided towards $\pi$ above ${\sqrt{\sm}=1.52\GeV}$ as only the $\kappa$ resonance below the $\keta$ threshold remains to be described by the input phase. 
The final results 
do not depend on the exact value of the
matching energy $\sm$, as long as it is chosen
in this range.
The resulting input phase shifts for the isospin-${1}/{2}$ and -${3}/{2}$ components are shown in Fig.~\ref{fig:Input_phase} as well. 

\section{Fit to scattering data}
\label{sec:fit_scattering}

We aim at a description of the scattering data from the $\kpi$ threshold up to $2.5\GeV$. In this energy range the 
particle data group (PDG) reports, 
besides the $\kappa$, two more resonances in the $S$-wave, $\Kb$ and $\Kc$~\cite{Zyla:2020zbs}.  We thus allow for two resonances in the resonance potential. 
Using a two-channel setup, incorporating the $\kpi$ and $\ketap$ channels, the model has a total of 6 free real parameters in $\VR$: 4 coupling constants and 2 masses. Following Ref.~\cite{Buettiker:2003pp}, we assume that the $\keta$ channel effectively decouples from $\kpi$. 
This assumption is confirmed by the analysis of Ref.~\cite{Pelaez:2016tgi}, which finds the $\kpi$ system elastic up to $1.6\GeV$. Moreover, we checked that an inclusion of the $\keta$ channel yields no significant difference of our results: the largest relative difference between a fit using a two-channel and three-channel model is about 0.5\% for the argument and 0.9\% for the modulus. Furthermore, the fit finds values consistent with zero for the couplings of the resonances to the $\keta$ channel. 

\begin{figure}[tp]
	\includegraphics[width=\linewidth,trim = 30pt 30pt 0pt 0pt]{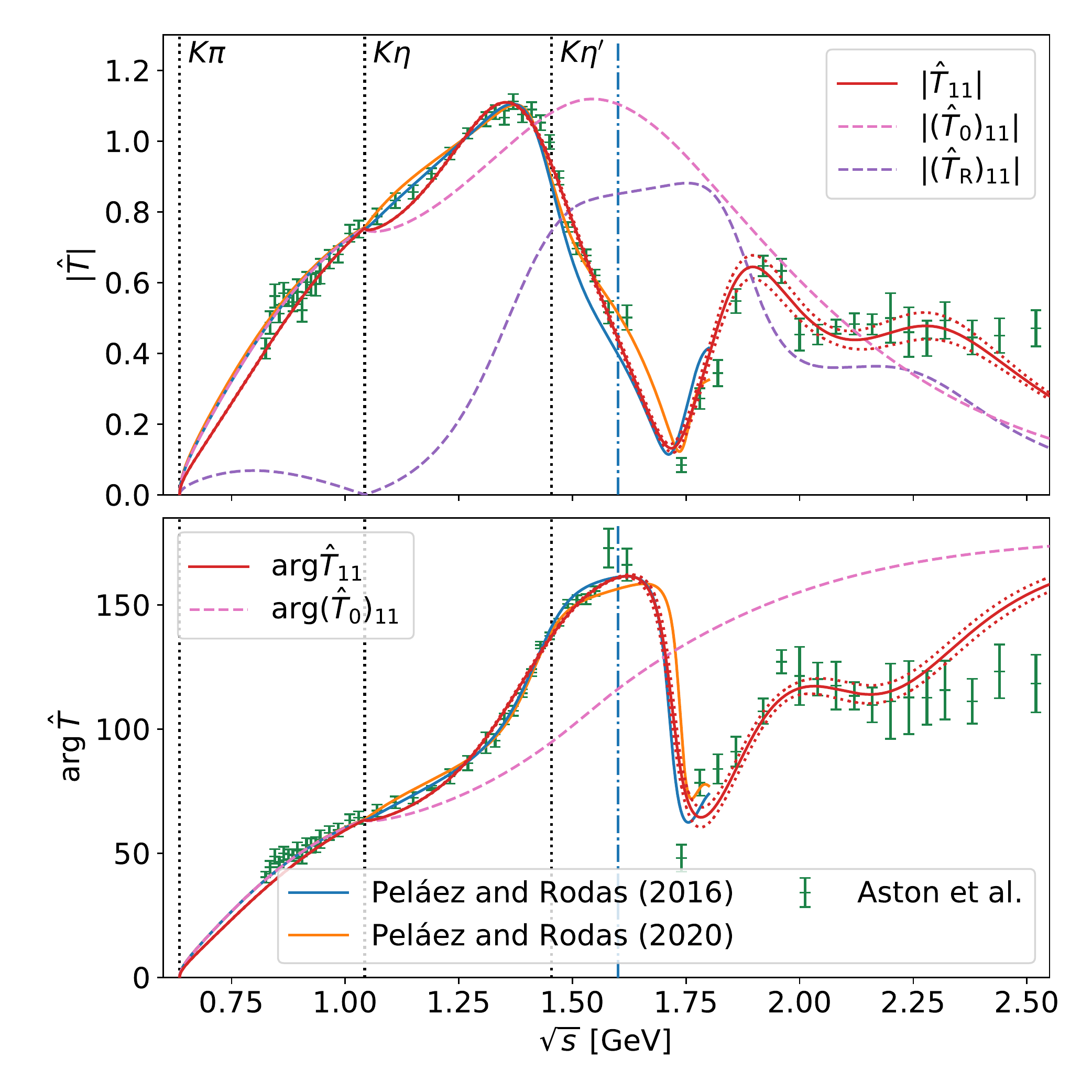}
	\centering
	\caption{Results for the combined fit of argument and absolute value of $\hat{T}$, defined in Eq.~\eqref{eq:That}, with 1$\sigma$ uncertainty band to the corresponding data of Aston et al.~\cite{Aston:1987ir}. We furthermore show the results of Pel\'aez and Rodas~(2016)~\cite{Pelaez:2016tgi} for comparison, which by the authors are quoted to be valid up to the dash-dotted line at $1.6\GeV$, and the newer results of Pel\'aez and Rodas~(2020)~\cite{Pelaez:2020gnd}. We moreover show the low-energy amplitude ${(\hat{T}_0)_{if}=\rho_i  \big(T^{\frac{1}{2}}_0 + T^{\frac{3}{2}}_0/2\big)_{if}}$ and the resonance part of the model ${(\hat{T}_\text{R})_{if}=\rho_i \big(T^{\frac{1}{2}}_\text{R}\big)_{if}}$ independently.}
	\label{fig:Fit_Aston}
\end{figure}%

\begin{table}[tp]
\renewcommand{\arraystretch}{1.3}
	\begin{tabular}{lr}
		\toprule
		\multicolumn{1}{l}{Parameter} & 
		\multicolumn{1}{c}{Value}\\
		\midrule
		$g^{(1)}_1~[\text{GeV}]$ & $ 2.898(29)$\\
		$g^{(1)}_2~[\text{GeV}]$ & $ -0.25(35)$\\
		$g^{(2)}_1~[\text{GeV}]$ & $ 2.14(17)$\\
		$g^{(2)}_2~[\text{GeV}]$ & $ 7.70(64)$\\
		$\mr{(1)}~[\text{GeV}]$ & $ 1.5708(33)$\\
		$\mr{(2)}~[\text{GeV}]$ & $ 2.133(36)$\\
		\midrule
		$\# \text{data points}$ & 112\\
		$\# \text{variables}$ & 6 \\
		$\chi^2$ & 370.8 \\
		$\chi^2/\#\text{d.o.f.}$ & 3.50 \\
		\bottomrule
	\end{tabular}
	\centering
	\caption{Parameters of the combined fit of argument and absolute value of $\hat{T}$, defined in Eq.~\eqref{eq:That}, to the corresponding data of Ref.~\cite{Aston:1987ir} as shown in Fig.~\ref{fig:Fit_Aston}.}
	\label{tab:fit_para}
\renewcommand{\arraystretch}{1.0}
\end{table}%

Figure~\ref{fig:Fit_Aston} shows the result of the combined fit of argument and modulus to the data set of Ref.~\cite{Aston:1987ir}, with the corresponding parameters given in Table~\ref{tab:fit_para}. The model is able to reproduce the data well up to about $2.3\GeV$. The subtraction point of the potential $s_0$ is fixed to the $\keta$ threshold with ${s_0=(\mk+\meta)^2}$. This choice is supported by fits where $s_0$ was treated as a  free parameter. Using a subtraction at ${s_0=0}$ as in the $\pi\pi$ analyses of Refs.~\cite{Hanhart:2012wi,Ropertz:2018stk} turns out to be insufficient to dampen the low-energy contributions of the resonance potential, as the $\kpi$ threshold lies much higher than the $\pi\pi$ threshold. With our choice for $s_0$, however, the full result matches the low-energy input closely below the $\keta$ threshold, as it should. The fit demonstrates that the coupling of the $\Kb$ to the $\ketap$ channel is small and within errors
consistent with zero, while the $\Kc$ couples strongly to $\ketap$.

The resulting reduced $\chi^2$ of about 3.5 seems rather unsatisfactory. However, comparing the  data of Ref.~\cite{Aston:1987ir} to the results of other groups such as Ref.~\cite{Estabrooks:1977xe} reveals that there are large discrepancies between the different data sets. Especially in the low-energy regime up to the opening of the $\keta$ threshold, a lot of data points differ by multiple standard deviations between the two sets. A combined fit of argument and modulus to both data sets more than doubles the reduced $\chi^2$, strongly indicating that some systematic uncertainties are underestimated---see also the related discussion in Ref.~\cite{Pelaez:2016tgi}. Hence, considering the modest quality of the data the fit performs quite decently. 
One could also try to extend the model to higher energies by adding an additional $K_0^*$ resonance.
However, this would require reliable data up to even higher energies, while we already cover the energy ranges of processes
of interest  such as ${\tau\rightarrow K_S\pi \nu_\tau}$ and ${{B}\rightarrow  {J}/\psi \kpi}$.

\begin{figure}[tp]
	\includegraphics[width=\linewidth,trim = 30pt 30pt 0pt 0pt]{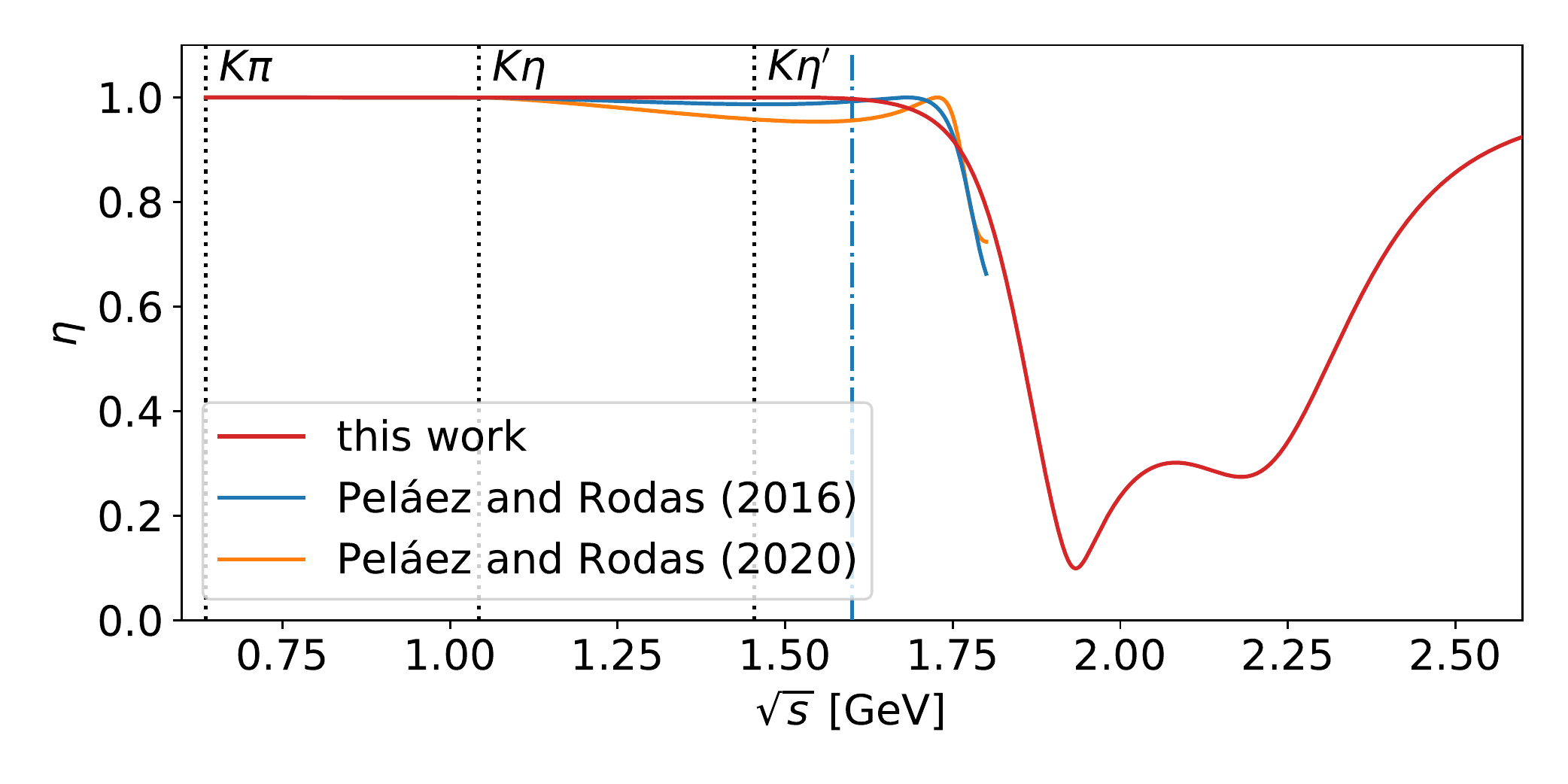}
	\centering
	\caption{Elasticities $\eta$ of the $\kpi$ isospin-$\frac{1}{2}$ wave extracted from Pel\'aez and Rodas~(2016)~\cite{Pelaez:2016tgi}, Pel\'aez and Rodas~(2020)~\cite{Pelaez:2020gnd}, and our model. The dashed blue line denotes the end of the range of validity of the analysis from Ref.~\cite{Pelaez:2016tgi} at $1.6\GeV$.}
	\label{fig:Eta}
\end{figure}

Figure~\ref{fig:Eta} shows the elasticity $\eta$ 
of the isospin-${1}/{2}$ amplitude that results
from the fit,   compared to that of the analysis of Ref.~\cite{Pelaez:2016tgi}.
One sees that our model is purely elastic up to $1.5\GeV$.  At higher energies, $\eta$ starts to decrease in a way consistent with Ref.~\cite{Pelaez:2016tgi}, although some deviations become visible.

\section{Application to $\tau$ decays}
\label{sec:tau}

As an application of the parameterization of the scalar form factor constructed based on the scattering input fixed in the preceding sections, we now focus on 
the reaction ${\tau^- \rightarrow \K_S \pi^- \nu_\tau}$, to improve the description of the spectrum 
measured by the Belle collaboration~\cite{Epifanov:2007rf} in the energy region where inelastic effects in the scalar form factor become relevant. In particular, we will study to which extent the excited $S$- and $P$-wave resonances $\Kb$ and $\Ke$ can be separated and provide an improved estimate of the $CP$ asymmetry produced by a tensor operator.

\subsection{Decay rate and form factor parameterization}

The differential decay rate can be parameterized by
\begin{align}
\frac{\text{d}\Gamma}{\text{d}\sqrt{s}} &=  \frac{c_\Gamma}{s}  \left(1 -\frac{s}{\mtau^2}\right)^2  \left(1 + 2 \frac{s}{\mtau^2}\right) q_{\kpi} \notag\\
& \qquad \times \left( q_{\kpi}^2 |\bar{f}_+|^2 + \frac{3 \Delta_{\kpi}^2}{4 s (1 + 2 \frac{s}{\mtau^2})} |\bar{f}_0|^2\right)\,,
\label{eq:BelleDecaywidth}
\end{align}
where ${\Delta_{\kpi}=\mk^2 - \mpi^2}$, the prefactor is given by
\begin{equation}
    c_\Gamma = \frac{G_F^2 \mtau^3}{96 \pi^3} S_\text{EW}^\tau \big(|V_{us}|f_+(0)\big)^2 \big(1 + \delta^{K\tau}_\text{EM}\big)^2\,, 
    \label{eq:cGamma}
\end{equation}
with the constants listed in Table~\ref{tab:Gamma_constants}, and 
\begin{equation}
    q_{\pi K}=\frac{\lambda^{1/2}\qty(s,\mpi^2,\mk^2)}{2\sqrt{s}}
\end{equation}
is the center-of-mass momentum of the $\pi K$ pair.

\begin{table}[tp]
\renewcommand{\arraystretch}{1.3}
	\begin{tabular}{lrr}
		\toprule
		\multicolumn{1}{l}{Quantity} & 
		\multicolumn{1}{c}{Value} & 
		\multicolumn{1}{c}{Reference}\\
		\midrule
		$G_F~ [10^{-5}\GeV^{-2}]$ & $1.1663787(6)$ & \cite{Tishchenko:2012ie}\\
		$S_\text{EW}^\tau$ & $1.0194$ & \cite{Marciano:1985pd,Marciano:1988vm,Braaten:1990ef}\\
		$|V_{us}|f_+(0)$ & $0.2165(4)$ & \cite{Zyla:2020zbs,Antonelli:2009ws}\\
		$\delta^{K\tau}_\text{EM}$ & $-0.15(20)\%$ & \cite{Antonelli:2013usa}\\
		\bottomrule
	\end{tabular}
	\centering
	\caption{Input quantities entering Eq.~\eqref{eq:cGamma}.}
	\label{tab:Gamma_constants}
	\renewcommand{\arraystretch}{1.0}
\end{table}

The actually measured events $N$ in an experimental setting in a bin at $\sqrt{s}$ then emerge from the decay rate as
\begin{equation}
    N = c_N \frac{\text{d}\Gamma}{\text{d}\sqrt{s}} := \frac{\lambda}{c_\Gamma} \frac{\text{d}\Gamma}{\text{d}\sqrt{s}}\,,
    \label{eq:BelleDecayevents}
\end{equation}
with $c_N$ some constant depending on the experimental setup. Here we assume the experimental binning to be chosen in such a way that the differential decay rate can be considered constant with respect to its uncertainty within one bin. For simplicity we combine all prefactors in the fits and define ${\lambda = c_N \times c_\Gamma}$, which remains a free parameter of the fit.

In the parameterization~\eqref{eq:BelleDecaywidth} the form factors are defined by the matrix elements
\begin{align}
 \langle\bar K^0(\pk)\pi^-(\ppi)|\bar s\gamma^\mu u|0\rangle&=(\pk-\ppi)^\mu f_+(s)\notag\\
 & \qquad +(\pk+\ppi)^\mu f_-(s)\,,\notag\\
 \langle\bar K^0(\pk)\pi^-(\ppi)|\bar s u|0\rangle&=\frac{\Delta_{\kpi}}{m_s-m_u}f_0(s)\,,
\end{align}
where
\beq
f_-(s)=\frac{\Delta_{\kpi}}{s}\big(f_0(s)-f_+(s)\big)\,.
\eeq
The Ward identity ensures the common normalization ${f_+(0)=f_0(0)}$ of vector and scalar form factors $f_+(s)$ and $f_0(s)$, which has been removed in the reduced form
\begin{equation}
   \bar{f}_+(s)=\frac{f_+(s)}{f_+(0)}\,,\qquad  \bar{f}_0(s)=\frac{f_0(s)}{f_+(0)}\,.
\end{equation}
With $\kpi$ $S$-wave scattering fixed as discussed in the previous section, the scalar form factor can be calculated via Eq.~\eqref{eq:FF} as ${\bar{f}_0(s)=(\FF)_1}$.
In principle, the vector form factor could also be described in a similar formalism, but for the present application we will employ a conventional parameterization from RChT~\cite{Moussallam:2007qc,Boito:2008fq,Boito:2010me,Bernard:2011ae,Antonelli:2013usa}, whose phase serves as input for an Omn\`es representation with three subtractions
\begin{align}
	\bar{f}_+(s) &= \exp\Bigg[\lambda^\prime \frac{s}{\mpi^2} + \frac{1}{2}\left(\lambda^{\prime\prime} - {\lambda^\prime}^2\right)\left(\frac{s}{\mpi^2}\right)^2  \notag\\
	&  \qquad + \frac{s^3}{\pi} \int_{s_\text{th}}^\infty \frac{\text{d}z}{z^3} \frac{\delta_1(z)}{(z-s-i\epsilon)}\Bigg]\,.\label{eq:PWaveOmnes}
\end{align}
Here, one subtraction constant was fixed by ${\bar{f}_+(0)=1}$, and the other two are related to the slope parameters of the form factor, which can be determined independently from $K_{\ell 3}$ decays. 
We choose them to be fixed by the central values of the results from Ref.~\cite{Antonelli:2013usa}, ${\lambda^\prime=25.621(405)\times10^{-3}}$ and ${\lambda^{\prime\prime}=1.2221(183)\times10^{-3}}$. 
As these parameters were not readjusted to the $\tau$ decay studied here, they impose an additional constraint on the small-$s$ behavior of the form factors, to which the $\tau$ spectrum is less sensitive. We have checked that the sum rules of Ref.~\cite{Antonelli:2013usa} for $\lambda^\prime$, $\lambda^{\prime\prime}$, which depend on the $P$-wave fit parameters,  remain well fulfilled 
in the fit, but otherwise will not propagate the corresponding uncertainties further, given that our focus lies on the inelastic part of the $\tau$ spectrum. 

 The phase $\delta_1$ in Eq.~\eqref{eq:PWaveOmnes} is parameterized as ${\arg(\hat{f}_+)}$ with $\hat{f}_+$ a RChT model for the form factor in terms of two resonances $\Kd$ and $\Ke$ and a mixing parameter $\beta$, given as
\begin{align}
\label{f+RChT}
	\hat{f}_+(s) &= \frac{\mr{\Kd}^2 - \kappa_{\Kd} \widetilde{H}_{\kpi}(0) + \beta s}{D\qty(\mr{\Kd},\gammar{\Kd})} \notag\\
	& \qquad - \frac{\beta s}{D\qty(\mr{\Ke},\gammar{\Ke})}\,,
\end{align}
with
\begin{equation}
	D\qty(\mr{\text{R}},\gammar{\text{R}}) = \mr{\text{R}}^2 -s - \kappa_\text{R} \Re \widetilde{H}_{\kpi}(s) - i \mr{\text{R}} \Gamma_\text{R}(s)
\end{equation}
and
\begin{align}
	\Gamma_\text{R}(s) &= \gammar{\text{R}} \frac{s}{\mr{\text{R}}^2} \bigg(\frac{\rho_1(s)}{\rho_1(\mr{\text{R}}^2)}\bigg)^3\,,\notag  \\
	\kappa_R &= \frac{1}{64 \pi^2} \frac{\gammar{\text{R}}}{\mr{\text{R}}} \frac{3 F_\K F_\pi}{\big(\rho_1(\mr{\text{R}}^2)\big)^3}\,.
\end{align}
Further,
\begin{align}
\widetilde{H}_{\kpi}(s)&=H(s)-\frac{2 L^r_9 s}{3 F_\K F_\pi}
=\frac{s M^r(s,\mu) - L(s)}{F_\K F_\pi}
\end{align}
is the $\kpi$ loop function in chiral perturbation theory (ChPT) with $H(s)$ as defined in Ref.~\cite{Gasser:1984ux}, where the chiral scale $\mu$ was fixed to ${\mu=M_{\Kd^0}=895.55\MeV}$~\cite{Zyla:2020zbs}. Explicit expressions for $M^r(s,\mu)$ and $L(s)$ can be found in Ref.~\cite{Gasser:1984gg} as well as for ${\widetilde{H}_{\kpi}(0)=H_{\kpi}(0)}$ in Ref.~\cite{Gasser:1984ux}. Note that the mass $\mr{\text{R}}$ and width $\gammar{\text{R}}$ parameters are bare parameters and do not correspond to physical masses and widths. The parameters for $\Kd$ and $\Ke$ are initially set to the results of Ref.~\cite{Antonelli:2013usa}, but are then allowed to vary within $2 \sigma$ for $\mr{\Kd}$ and $\gammar{\Kd}$, $5 \sigma$ for $\mr{\Ke}$, $1.5\sigma$ for $\gammar{\Ke}$, and $10 \sigma$ for $\beta$, although ${\beta<0}$ is still enforced. These parameter ranges were chosen in such a way that the shape of the generated $\kpi$ $P$-wave scattering phase shift remains phenomenologically viable. 

Since we employ a two-channel formalism for the $S$-wave, the parameters to be adjusted to the $\tau$ decay data are the normalization constants $c_1^{(0)}$ and $c_2^{(0)}$, potentially to be supplemented by higher terms in the polynomial for the source term, as well as source--resonance couplings $\alpha^{(1)}$ and $\alpha^{(2)}$ to the two resonances. 
Due to the normalization ${\bar{f}_0(0)=1}$, given by the Ward identity, ${f_+(0)=f_0(0)}$, the constant term ${c_1^{(0)}}$ is implicitly fixed by ${\bar{f}_0(0)=(\FF)_1(0)=1}$. Furthermore the normalization of the $\ketap$ scalar form factor is fixed from matching to the corresponding expression from $U(3)$ ChPT at ${s=0}$, which, with the standard single $\eta$--$\eta^\prime$ mixing angle, is larger than the $\kpi$ scalar form factor by a factor of $\sqrt{3}$ at leading order, resulting in ${(\FF)_2(0)=M_2(0)=\sqrt{3}}$, which implicitly fixes $c_2^{(0)}$.
Higher-order corrections tend to reduce this result~\cite{Jamin:2001zq}, however, as we will find, the sensitivity of the data to the $\ketap$ channel is limited, so that the leading-order estimate is sufficient for our purposes. 

Higher polynomials in the source term have the potential to improve the description of the scalar form factor in the $\tau$ decay region, at the cost of changing its high-energy behavior. We therefore investigate the influence of a linear term in $s$ proportional to $c^{(1)}_1$ for the $\kpi$ channel. For the $\ketap$ channel on the other hand, this did not prove necessary as already the leading-order constant is poorly determined in the fit. Furthermore, our phase description of the $S$-wave does not only include the $\Kb$ resonance, which is perfectly within the decay region, but also the $\Kc$ resonance, which lies significantly above the $\tau$ mass. Hence it is to be expected that the corresponding source-term coupling $\alpha^{(2)}$ is difficult to constrain via the $\tau$ decay data. Accordingly, we will consider fit variants in which the $\Kc$ source-term coupling is set to zero for this decay. Note that this does not remove the $\Kc$ resonance completely from our model, as the phase still contains the full information about all resonances. This is a distinct feature of this construction, reflecting the built-in unitarity constraints. 

Finally, we introduce a further restriction into our fitting routine: the Callan--Treiman low-energy theorem~\cite{Callan:1966hu,Dashen:1969bh,Gasser:1984ux,Bijnens:2007xa,Kastner:2008ch} constrains the scalar $\kpi$ form factor below threshold at ${s=\Delta_{\kpi}}$ to
\begin{equation}
    \bar{f}_0 (\Delta_{\kpi}) = \frac{F_\K}{F_\pi} + \Delta_{CT}\,,
\end{equation}
where $\Delta_{CT}$ is a very small correction. To implement this condition we introduce an additional term to the $\chi^2$ sum weighted by $\Delta_{CT}$, given as 
\begin{equation}
    \chi^2 \rightarrow \chi^2 + \left(\frac{\bar{f}_0(\Delta_{\kpi}) - ({F_\K}/{F_\pi}) - \Delta_{CT}}{\Delta_{CT}}\right)^2\,.
    \label{eq:xi_additional}
\end{equation}
We take ${\Delta_{CT}=-5.6 \times 10^{-3}}$ from Ref.~\cite{Bijnens:2007xa}, which includes isospin breaking and corrections up to next-to-next-to-leading order. For the ratio of the decay constants we use ${F_K/F_\pi\sim 1.195}$~\cite{Aoki:2019cca}.

\subsection{Fit results}

\begin{table*}[tp]%
\renewcommand{\arraystretch}{1.3}
	\begin{tabular}{lrrrr}
		\toprule
		\multicolumn{1}{l}{Parameter}&\multicolumn{1}{c}{Fit~1} &\multicolumn{1}{c}{Fit~2} &\multicolumn{1}{c}{Fit~3}&\multicolumn{1}{c}{Fit~4}\\
		\midrule
		$\lambda$             & $0.753(11)$  & $0.7440(94)$  & $0.7617(87)$    & $0.746(11)$\\
		$\alpha^{(1)} ~[\text{GeV}]$         & $-0.35(26)$    & $-0.28(21)$   & $0.035(40)$   & $-0.42(19)$\\
		$\alpha^{(2)}~[\text{GeV}]$ & $1.9(1.3)$     & $-4.3(3.9)$    & $0$ (fixed)     & $0$ (fixed)\\
		$c^{(1)}_1~[\text{GeV}^{-2}]$   & $0$ (fixed)   & $-0.65(34)$   & $0$ (fixed)     & $-0.25(11)$\\
		\midrule
		$\mr{\Kd}~[\text{MeV}]$        & $943.71(57)$  & $943.26(53)$    & $944.04(52)$   & $943.40(54)$\\
		$\gammar{\Kd}~[\text{MeV}]$   & $67.15(88)$   & $66.46(82)$     & $67.61(80)$    & $66.69(82)$\\
		$\mr{\Ke}~[\text{MeV}]$        & $1355(34)$    & $1381(39)$      & $1354(15)$     & $1357(24)$\\
		$\gammar{\Ke}~[\text{MeV}]$   & $205(100)$     & $205(100)$       & $229(22)$      & $176(35)$\\
		$\beta$                     & $-0.032(16)$  & $-0.029(12)$    & $-0.0418(48)$  & $-0.0251(75)$\\
		\midrule
		$\# \text{data points}$ & 97+1 & 97+1 & 97+1 & 97+1 \\
		$\# \text{variables}$ & 8 & 9 & 7 & 8 \\
		$\chi^2$ & 93.1 & 87.4 & 97.7 & 89.4 \\
		$\chi^2/\#\text{d.o.f.}$ & 1.03 & 0.98 & 1.07 & 0.99 \\
		\bottomrule
	\end{tabular}
	\centering
	\caption{Parameters of the fits of theoretical events $N$, defined in Eq.~\eqref{eq:BelleDecayevents}, to  efficiency-corrected and background-reduced events for ${\tau^- \rightarrow K_S \pi^- \nu_\tau}$~\cite{Epifanov:2007rf} including the additional constraint of Eq.~\eqref{eq:xi_additional} with different combinations of fixed $\alpha^{(2)}$ and $c^{(1)}_1$ parameters. As the outcome of all fits is quite close, only Fit~3 is shown exemplarily in Fig.~\ref{fig:Fit_Decay}, while we display the comparison between the fits in Fig.~\ref{fig:Fit_Decay_rel}.}
	\label{tab:fit_decay_para}
\renewcommand{\arraystretch}{1.0}
\end{table*}%
\begin{figure}[tp]%
	\includegraphics[width=\linewidth,trim = 30pt 30pt 0pt 0pt]{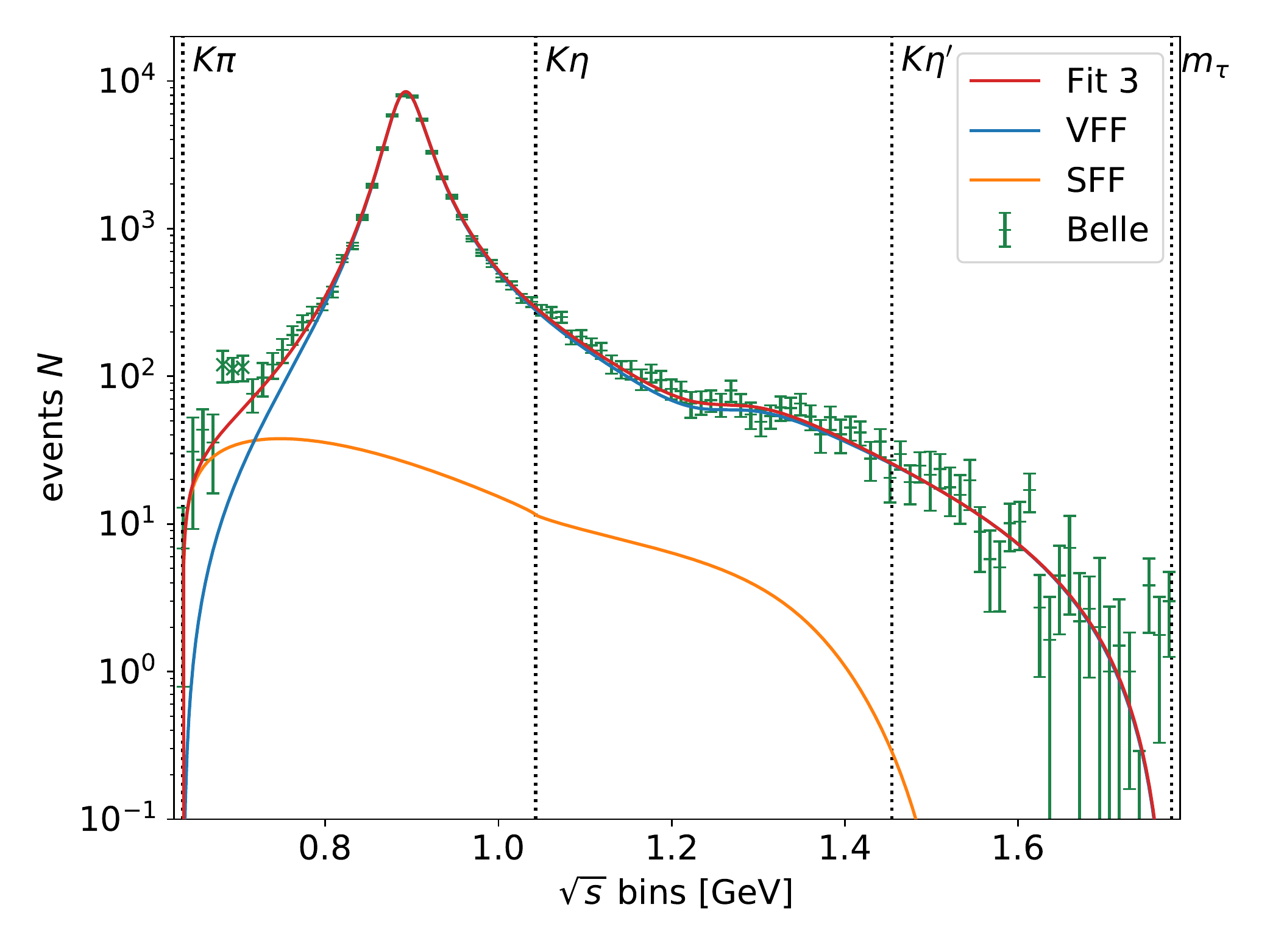}
	\centering
	\caption{Example result Fit~3 for the fit of theoretical events $N$, defined in Eq.~\eqref{eq:BelleDecayevents}, to efficiency-corrected and background-reduced events for ${\tau^- \rightarrow K_S \pi^- \nu_\tau}$ of Ref.~\cite{Epifanov:2007rf} including the additional constraint of Eq.~\eqref{eq:xi_additional}. In addition, we show the scalar form factor (SFF) and vector form factor (VFF) components separately.}
	\label{fig:Fit_Decay}
\end{figure}%
\begin{figure}[tp]%
	\includegraphics[width=\linewidth,trim = 30pt 30pt 0pt 0pt]{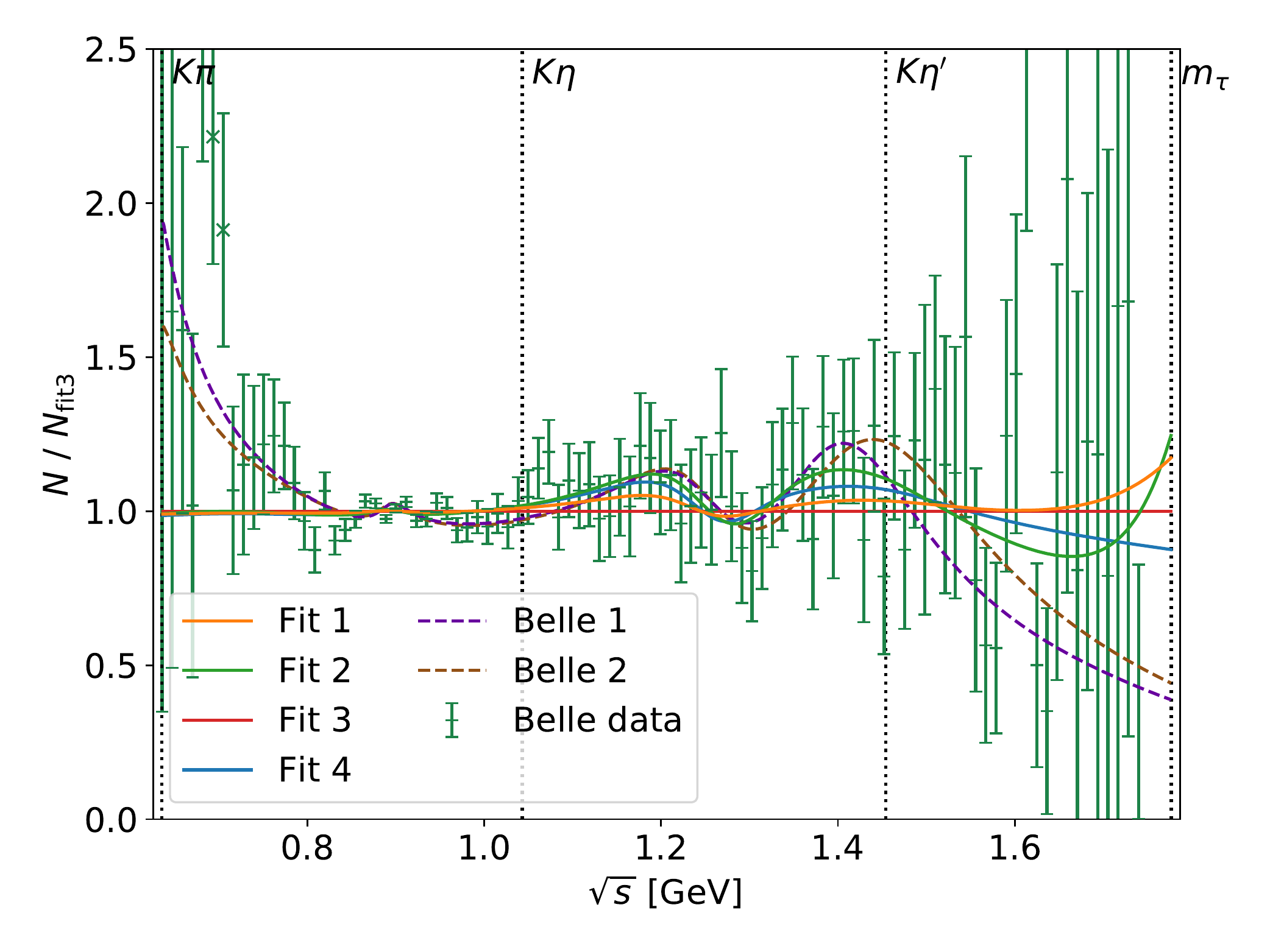}
	\centering
	\caption{Comparison of the different fit results presented in Table~\ref{tab:fit_decay_para} as well as two BW parameterizations ``Belle 1'' and ``Belle 2''~\cite{Epifanov:2007rf}---which include $\Ka$, $\Kd$, and $\Ke$, or $\Ka$, $\Kd$, and $\Kb$, respectively---all normalized by Fit~3.}
	\label{fig:Fit_Decay_rel}
\end{figure}%

We consider the four fit variants presented in Table~\ref{tab:fit_decay_para}. 
As indicated, we distinguish between fits with and without the $\Kc$ source-term coupling $\alpha^{(2)}$ as well as with and without a linear term in $s$ proportional to $c^{(1)}_1$ in the source term of the scalar form factor. The parameters of the scalar resonances are kept fixed to their values
determined in the fit to the scattering data. 
As all fits are in agreement with each other and of similar quality, Fig.~\ref{fig:Fit_Decay} shows the results only of Fit~3, together with the efficiency-corrected and background-subtracted events as measured by Belle~\cite{Epifanov:2007rf} as well as the separate contributions from the vector and scalar form factor, respectively. 
For all fits we excluded the data points 5, 6, and 7, following Refs.~\cite{Jamin:2008qg,Boito:2008fq,Boito:2010me,Antonelli:2013usa}. The inclusion of these points would increase the $\chi^2/\#\text{d.o.f.}$ by $0.15$--$0.2$ without any significant shift in the fit parameters, suggesting a conflict with the general principles on which our fit function is based. Since the experimental uncertainties included in the fit (and shown in Fig.~\ref{fig:Fit_Decay}) are only statistical, this is likely due to an unaccounted-for systematic effect.    
The relative differences between the various fits as well as the comparison to two of the original Belle BW parameterizations~\cite{Epifanov:2007rf} are displayed in Fig.~\ref{fig:Fit_Decay_rel}, which are normalized to the result of Fit~3. ``Belle 1'' corresponds to a BW description including $\Ka$, $\Kd$, and $\Ke$, and ``Belle 2'' contains $\Ka$, $\Kd$, and $\Kb$. With their BW framework, Belle was only able to describe the structure around $1.4\GeV$ either by the vector $\Ke$ or by the scalar $\Kb$ resonance, but not by both at the same time.

The $\tau$ decay spectrum is highly dominated by the vector form factor and the $\Kd$ resonance, making other components of the decay rate difficult to separate. However, it is known that a description in terms of the $\Kd$ resonance alone is not sufficient, as also found by Belle~\cite{Epifanov:2007rf}. In our analysis, we find meaningful fits of the decay spectrum, despite the strong overlap  between the $\Kb$ and $\Ke$ resonances.
With information on the $\Kb$ resonance entering via the $\pi K$ $S$-wave phase shift, the fit to the spectrum allows us to determine the mass parameter of the $\Ke$ at the level of $30\MeV$, so that the combination of scattering data and the $\tau$ spectrum permits some discrimination between the $S$- and $P$-wave resonances even without additional 
differential information (see below).
As expected, the influence of the $\Kc$ resonance in the decay region is very small and the fit results with and without $\Kc$ source-term coupling are nearly indistinguishable in terms of the decay spectrum, as reflected by the large uncertainties on $\alpha^{(2)}$ in Fits~1 and 2. The linear term in the source term $c^{(1)}_1$, on the other hand, improves the fit more substantially and does not come out consistent with zero. However, as the fit quality is already sufficient without it, we cannot claim conclusive evidence for the necessity of a linear term either. At the current level of precision, we thus conclude that the four fit variants are essentially equivalent. 

\begin{figure}[tp]%
	\includegraphics[width=\linewidth,trim = 30pt 30pt 0pt 0pt]{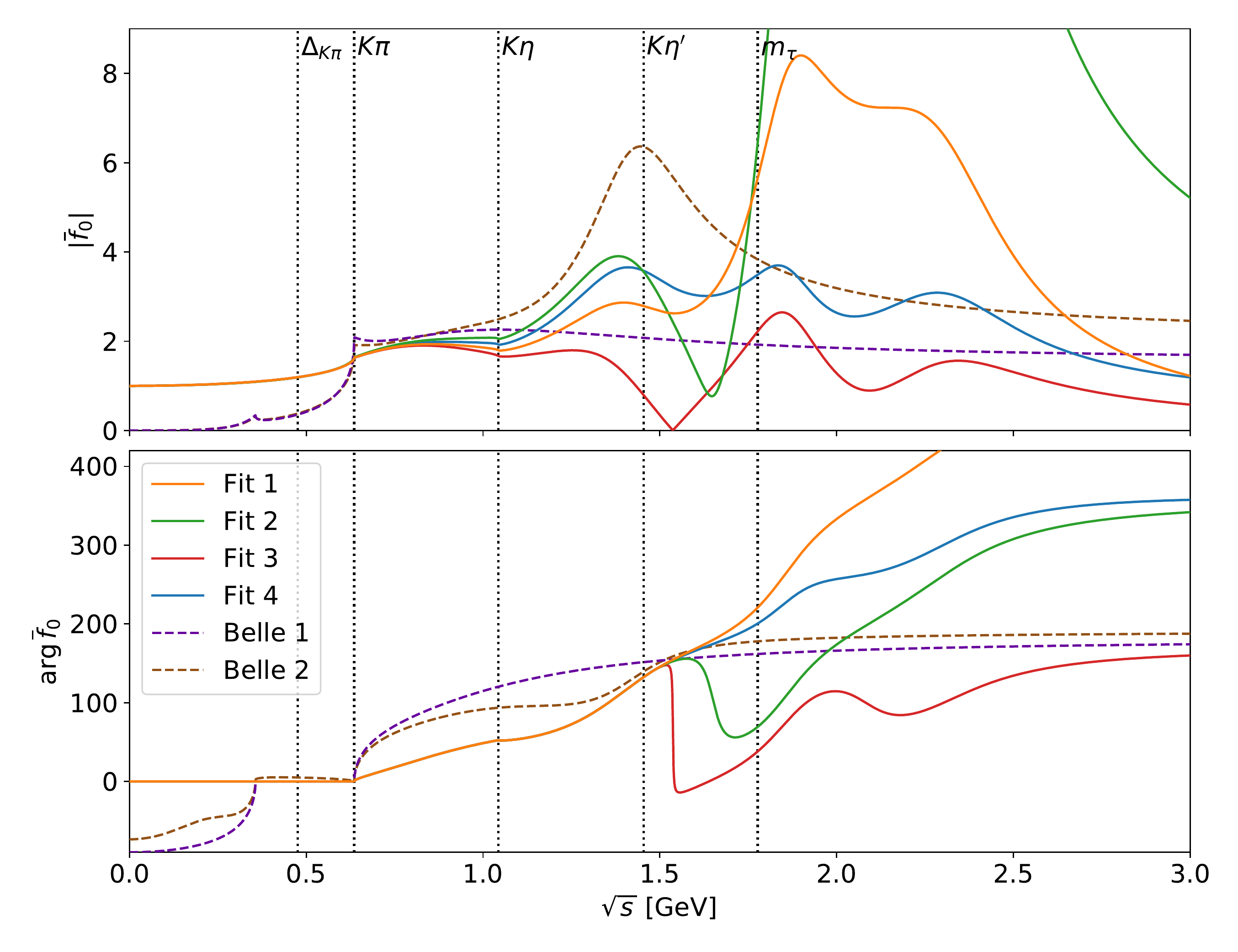}
	\centering
	\caption{Scalar form factor $\bar{f}_0$ (top: modulus, bottom: phase) with the parameters of the fit results of Table~\ref{tab:fit_decay_para}.}
	\label{fig:Fit_SFF}
\end{figure}%
\begin{figure}[tp]%
	\includegraphics[width=\linewidth,trim = 30pt 30pt 0pt 0pt]{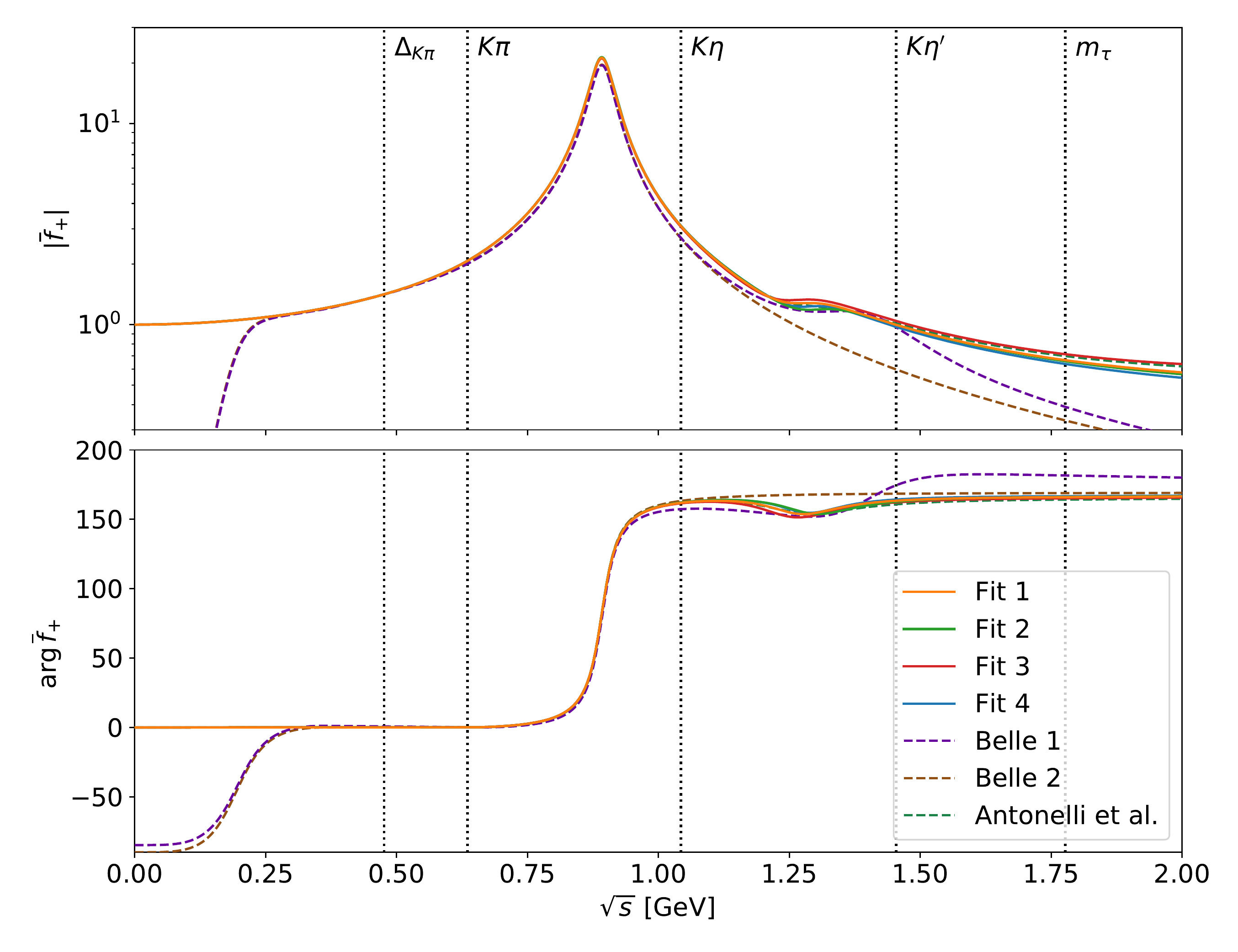}
	\centering
	\caption{Vector form factor $\bar{f}_+$  (top: modulus, bottom: phase) with the parameters of the fit results of Table~\ref{tab:fit_decay_para}.}
	\label{fig:Fit_VFF}
\end{figure}%

Examining the underlying scalar and vector form factors, as shown in Figs.~\ref{fig:Fit_SFF}  and~\ref{fig:Fit_VFF}, respectively, the advantages of our parameterization in comparison to the BW approach become evident. 
By construction, the phase of the scalar form factor coincides with the scattering phase up to the $\ketap$ threshold. As only the absolute value of the form factor enters into Eq.~\eqref{eq:BelleDecaywidth}, the measurement cannot fix its phase directly, but it is determined implicitly in accord with the unitarity condition, a constraint clearly violated by the BW parameterizations, see Fig.~\ref{fig:Fit_SFF}. 
Furthermore, contrary to the BW model, our representation fulfills the Callan--Treiman low-energy theorem up to at least 0.5\%, which corresponds to less than 10\% of $\Delta_\text{CT}$. 

The vector form factor comes out close to the results of Ref.~\cite{Antonelli:2013usa}, from where its parameterization originates, with small differences in the inelastic region due to the use of our improved parameterization of the scalar form factor. 
The constraints on $\lambda^\prime$ and $\lambda^{\prime\prime}$ using the $K_{\ell3}$ input from Ref.~\cite{Antonelli:2013usa} are still fulfilled up to at least 0.5\%. 
The result is also relatively close to the BW parameterization, which works well as long as the $\Kd$ resonance dominates. However, unitarity violation still occurs in the threshold region due to unphysical imaginary parts, and the phase differs considerably as soon as the $\Ke$ resonance becomes relevant. 

Comparing the four fits, differences emerge starting around the $\ketap$ threshold. The phase of Fit~3 largely follows the elastic input phase, in Fit~2 still a sharp drop-off occurs, while in Fits~1 and 4 no such effect is visible. This behavior is mirrored in the modulus, almost reaching zero in Fit~3 and a pronounced minimum in Fit~2. Further,       
the results of Fits~1 and 3, which do  not involve a slope parameter $c^{(1)}_1$, tend to have a smaller scalar form factor in the $\tau$ decay region and a slightly lower bare mass for the $\Ke$ in the vector form factor. On the other hand, Fits~2 and 4, including a slope $c^{(1)}_1$, have more freedom to increase the scalar form factor at lower energies, which results in a slightly higher $\Ke$ bare mass and a smaller value of $\lambda$. Asymptotically, the scalar form factors without slope fall off like $1/s$ for high energies, as expected from perturbative QCD~\cite{Lepage:1979zb,Lepage:1980fj}, while those with a slope approach a constant; again, the $\tau$ spectrum is not sufficient to differentiate. 
In fact, the scalar form factors beyond the $\ketap$ threshold are not well constrained at all, as that region is already strongly suppressed by phase space in the decay spectrum and the data points have large uncertainties. This is the reason why Fits~3 and 4 are much better behaved when extrapolated beyond the energy region probed in the $\tau$ decay, since without setting ${\alpha^{(2)}=0}$ the fit function can extend to large values before the asymptotic behavior sets in. 
Finally, one finds that all scalar form factors still generate resonant structures above the $\tau$ decay region, even if the source-term couplings are set to zero: as we already remarked above, unitarity demands that the underlying phase still contain information about all resonances.

\begin{figure}[tp]%
	\includegraphics[width=\linewidth,trim = 30pt 30pt 0pt 0pt]{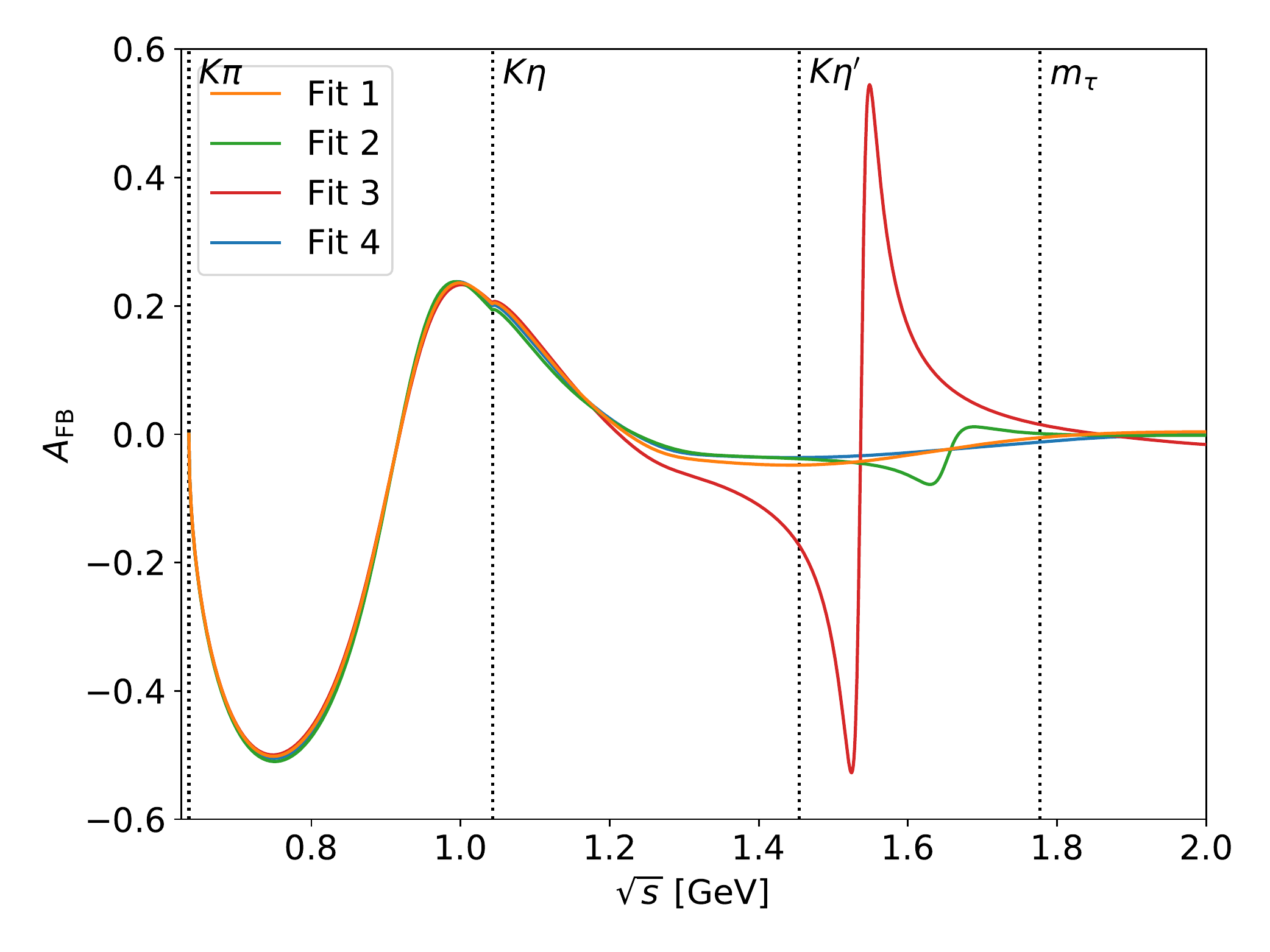}
	\centering
	\caption{Forward--backward asymmetry as defined in Eq.~\eqref{eq:A_FB} for the four fit results as given in Table~\ref{tab:fit_decay_para}.}
	\label{fig:AsymFB}
\end{figure}

Since the scalar resonance $\Kb$ and the vector resonance $\Ke$ occupy the same energy region, ultimately additional data beyond the spectrum are required to better determine their parameters. One such observable 
that separates vector and scalar components is the forward--backward asymmetry~\cite{Beldjoudi:1994hi,Kou:2018nap}
\begin{align}\nonumber
A_\text{FB}(s)&= \frac{\int_0^1 \text{d}z\left[\frac{\text{d}\Gamma}{\text{d}z}(z)-\frac{\text{d}\Gamma}{\text{d}z}(-z)\right]}{\int_0^1 \text{d}z\left[\frac{\text{d}\Gamma}{\text{d}z}(z)+\frac{\text{d}\Gamma}{\text{d}z}(-z)\right]} \\
&=\frac{- 2 \Re(f_0 f_+^*) \Delta_{\kpi} q_{\kpi} \sqrt{s}}{|f_0|^2 \Delta_{\kpi}^2 + \frac{4}{3} |f_+|^2 q_{\kpi}^2 \qty(\frac{2 s^2}{\mtau^2} + s)}\,,\label{eq:A_FB}
\end{align}
where $z$ denotes the cosine of the $\pi K$ helicity angle.
The quantity $A_\text{FB}(s)$ can potentially be measured at Belle II~\cite{Kou:2018nap}.
We show the predictions corresponding to the four fits in Fig.~\ref{fig:AsymFB}. As expected, the different fits are quite distinct above the $\ketap$ threshold
due to the different phase motion, allowing one to distinguish among them once data on $A_\text{FB}(s)$ become available.

Finally, by integrating over Eq.~\eqref{eq:BelleDecaywidth} and normalizing with respect to the total $\tau$ decay width ${\Gamma_\tau=2.267(4)\times 10^{-3}\,\text{eV}}$~\cite{Zyla:2020zbs}, we can calculate the branching ratio of ${\tau\rightarrow K_S \pi \nu_\tau}$ for the different fits:
\begin{align}
    \text{BR}({\tau\rightarrow K_S \pi \nu_\tau})\Big|_\text{Fit~1}
    &= 4.334(66)(25)\times10^{-3}\,,\notag\\
    \text{BR}({\tau\rightarrow K_S \pi \nu_\tau})\Big|_\text{Fit~2}
    &= 4.390(48)(26)\times10^{-3}\,,\notag\\
    \text{BR}({\tau\rightarrow K_S \pi \nu_\tau})\Big|_\text{Fit~3}
    &= 4.284(35)(25)\times10^{-3}\,,\notag\\
    \text{BR}({\tau\rightarrow K_S \pi \nu_\tau})\Big|_\text{Fit~4}
    &= 4.377(49)(26)\times10^{-3}\,,
\end{align}
where the first error refers to the statistical uncertainty propagated from the fit parameters and the second one to the normalization, see Table~\ref{tab:Gamma_constants}. In this way, we do not make an attempt to extract the absolute normalization from the $\tau$ data, but rather perform a consistency check with the $K_{\ell 3}$ data. Moreover, we do not propagate the systematic uncertainties incurred indirectly when using $K_{\ell3}$ input for the subtraction constants $\lambda^\prime$ and $\lambda^{\prime\prime}$, nor
the uncertainties on the $S$-wave phase as given in Table~\ref{tab:fit_para}, which are expected to be negligible in this application as they only become relevant above the $\tau$ mass.

Since we cannot give preference to any particular fit variant, we quote the average over all four versions as central value and assign the spread as systematic uncertainty
\begin{align}
    \text{BR}({\tau\rightarrow K_S \pi \nu_\tau}) &=
    4.35(6)_\text{st}(3)_\text{norm}(7)_\text{sys}\times10^{-3}\notag\\
    &=4.35(10)\times10^{-3}\,.
\end{align}
This result lies $2\sigma$ above the original Belle result $\text{BR}({\tau\rightarrow K_S \pi \nu_\tau})|_\text{\cite{{Epifanov:2007rf}}}=4.04(13)\times 10^{-3}$, but agrees with the more recent $\text{BR}({\tau\rightarrow K_S \pi \nu_\tau})|_\text{\cite{{Ryu:2014vpc}}}=4.16(8)\times 10^{-3}$ as well as the PDG average $\text{BR}({\tau\rightarrow K_S \pi \nu_\tau})|_\text{\cite{{Zyla:2020zbs}}}=4.19(7)\times 10^{-3}$ at the level of $1.5\sigma$.
We thus conclude that the branching fraction extracted by combining the shape as measured in the Belle spectrum with the normalization from $K_{\ell 3}$ decays comes out consistent with the direct measurement in $\tau$ decays.

\subsection{$CP$ asymmetry}

The $CP$ asymmetry in $\tau\to K_S\pi\nu_\tau$ is defined as
\beq
 A_{CP}^\tau=\frac{\Gamma(\tau^+\to\pi^+ K_S\bar \nu_\tau)-\Gamma(\tau^-\to\pi^- K_S \nu_\tau)}{\Gamma(\tau^+\to\pi^+ K_S\bar \nu_\tau)+\Gamma(\tau^-\to\pi^- K_S \nu_\tau)}\,.
\eeq
In the Standard Model, it is dominated by indirect $CP$ violation, leading to the prediction~\cite{Zyla:2020zbs}
\begin{align}
 A_{CP}^\tau&=A_L=\frac{\Gamma(K_L\to\pi^-\ell^+\nu_\ell)-\Gamma(K_L\to\pi^+\ell^-\bar\nu_\ell)}{\Gamma(K_L\to\pi^-\ell^+\nu_\ell)+\Gamma(K_L\to\pi^+\ell^-\bar\nu_\ell)}\notag\\
 &=3.32(6)\times 10^{-3}\,,
\end{align}
which is in conflict with the 2012 measurement by the BaBar collaboration~\cite{BABAR:2011aa}
\beq
\label{CP_tau_exp}
A_{CP}^{\tau,\text{exp}}=-3.6(2.3)(1.1)\times 10^{-3}\,.
\eeq
Including small corrections related to the $K_S$ reconstruction~\cite{Grossman:2011zk}, this amounts to a $2.8\sigma$ tension. 

As pointed out in Refs.~\cite{Devi:2013gya,Dhargyal:2016jgo}, due to the absence of a scalar--vector interference there are limited options to produce an effect with physics beyond the Standard Model (BSM), the only remaining option being a tensor--vector interference. Estimates for its size then depend on the tensor form factor defined by 
\begin{equation}
    \langle\bar K^0(\pk)\pi^-(\ppi)|\bar s\sigma^{\mu\nu} u|0\rangle=i\frac{\pk^\mu\ppi^\nu-\pk^\nu\ppi^\mu}{\mk}B_T(s)\,,\!\!
\end{equation}
based on which the $CP$ asymmetry takes the form~\cite{Cirigliano:2017tqn}
\begin{align}
\label{CP_BSM}
A_{CP}^{\tau,\text{BSM}}&=\frac{\Im c_T}{\Gamma_\tau\text{BR}(\tau\to K_S\pi\nu_\tau)}\\
&\hspace{-1cm}\times\int_{s_{\pi K}}^{\mtau^2}\text{d}s'\kappa(s')|f_+(s')||B_T(s')|
\sin\big(\delta_+(s')-\delta_T(s')\big)\,,\notag
\end{align}
where ${s_{\pi K}=(\mpi+\mk)^2}$, $\delta_T(s)$ is the phase of $B_T(s)$,
\begin{equation}
    \kappa(s)=G_F^2|V_{us}|^2S_\text{EW}^\tau\frac{q_{\kpi}^3(\mtau^2-s)^2}{32\pi^3\mtau^2\mk \sqrt{s}}\,,
\end{equation}
and $\Im c_T$ is the imaginary part of the tensor Wilson coefficient. The key observation made in Ref.~\cite{Cirigliano:2017tqn} is that Watson's theorem implies ${\delta_+(s)=\delta_T(s)}$, so that the strong phase due to the $\Kd$ cancels in Eq.~\eqref{CP_BSM}, with the remaining inelastic effect due to the $\Ke$ suppressed by two orders of magnitude. A simple BW estimate was used to argue
\beq
\label{ACP_cT}
\big|A_{CP}^{\tau,\text{BSM}}\big|_\text{\cite{Cirigliano:2017tqn}}\lesssim 0.03 |\Im c_T|\,,
\eeq
which, together with limits on $\Im c_T$ from the neutron electric dipole moment and $D$--$\bar D$ mixing, was sufficient to exclude this explanation of the BaBar measurement~\eqref{CP_tau_exp}.

The estimate~\eqref{ACP_cT} was subsequently revisited in Refs.~\cite{Rendon:2019awg,Dighe:2019odu,Chen:2019vbr}, mostly in the framework of RChT, including the suggestion in Ref.~\cite{Chen:2019vbr} to constrain the tensor form factor using large-$N_c$ arguments~\cite{Cata:2008zc}. As another application of the improved treatment of the scalar form factor and, in consequence, the vector form factor as extracted from the $\tau$ spectrum, we now turn to a refined evaluation of Eq.~\eqref{CP_BSM}. 

While the normalization is determined from lattice QCD, ${B_T(0)=0.686(25)}$~\cite{Baum:2011rm}, 
to constrain the phase of the tensor form factor beyond the elastic region, we need information about the coupling of the $\Ke$ to the tensor current relative to its vector-current coupling, which in the parameterization~\eqref{f+RChT} is contained in $\beta$. 
As pointed out in Ref.~\cite{Chen:2019vbr}, such a constraint can be extracted from 
the large-$N_c$ pattern derived in Ref.~\cite{Cata:2008zc},%
\beq
\xi_n=\frac{f_{V_n}^{T}}{f_{V_n}}\to (-1)^n \frac{1}{\sqrt{2}}\,,
\eeq
for the ratio of tensor over vector coupling constants for the $n$th excitation of a vector meson. For the ground state ${n=0}$ one has~\cite{Hoferichter:2018zwu}
\beq
\label{fn0}
\frac{f_{\Kd}^{T}}{f_{\Kd}}=B_T(0)\frac{M_{\Kd}}{2\mk}\sim 0.62\,,
\eeq
indeed close to $1/\sqrt{2}$. Denoting the tensor analog of $\beta$ by $\gamma$, these arguments lead to
\beq
\gamma = -\frac{M_{\Kd}}{M_{\Ke}}\beta\sim -0.63 \beta\,,
\eeq
which would change to
\beq
\gamma = - \frac{\sqrt{2} \mk}{B_T(0) M_{\Ke}}\beta\sim -0.73\beta 
\eeq
if the lattice-QCD number~\eqref{fn0} were used for the $\Kd$, but the asymptotic value $1/\sqrt{2}$ for the $\Ke$. In the following, we will thus use the estimate ${\gamma=-0.7\beta}$, discarding the (unlikely) possibility of the opposite sign~\cite{Cata:2008zc}, in which case the $CP$ asymmetry would be even further suppressed. 

Averaging over the four fit variants given in Table~\ref{tab:fit_decay_para}, with ${\beta\to\gamma}$ for the tensor form factor, we obtain
\begin{align}
\label{CPasym}
 A_{CP}^{\tau,\text{BSM}}&=-0.034(11)(7)(5)\,\Im c_T\notag\\
 &=-0.034(14)\,\Im c_T\,,
\end{align}
where the uncertainties refer to the systematic effects when including $c_1^{(1)}$, $\alpha^{(2)}$, and a $30\%$ large-$N_c$ uncertainty assigned to the tensor phase, respectively. 
The final result thus nicely confirms the simple estimate of Eq.~\eqref{ACP_cT}. 

\section{Pole extraction}
\label{sec:poles}

Resonances manifest themselves as poles on the unphysical Riemann sheets of the $S$- or, equivalently, the $T$-matrix. The pole position in the complex plane, $s_\text{R}$,
is conventionally parameterized in terms of a mass parameter $M_\text{R}$ and a width parameter $\Gamma_\text{R}$
via
\begin{equation}
	\sqrt{s_\text{R}}=M_\text{R}-i \frac{\Gamma_\text{R}}{2}\,.
	\label{eq:pole_defintion}
\end{equation}
For resonances distorted by threshold effects or overlapping resonances, the
value of $\Gamma_\text{R}$ neither agrees with the visible width nor can it be directly related to
the lifetime of the state.
Moreover, for these cases and broad resonances, $M_\text{R}$ and $\Gamma_\text{R}$
can deviate significantly from the corresponding  BW parameters, which are model- and reaction-dependent quantities.

As $T_0$ has a complicated analytic structure due to left-hand cuts, which cannot be deduced from the phase shift alone, an analytic continuation to other Riemann sheets is not feasible. However, as $T$-matrix 
and form factor are smooth functions when moving across a cut from the physical Riemann sheet to the connected
unphysical sheet, we can use Pad\'e approximants to determine the nearest pole on the neighboring unphysical sheet.

Assuming that an amplitude $F(s)$ is analytic inside the disc $B_\delta(s_0)$ around some expansion point $s_0^{(N)}$
except for one simple pole, we can expand $F(s)$ according to Montessus' theorem as
\begin{equation}
	P^N_1(s,s_0)=\frac{ \sum_{n=0}^{N} a_n^{(N)} \qty(s-s_0^{(N)})^n}{1+ b^{(N)} \qty(s-s_0^{(N)})}\,,
	\label{eq:pade_definition}
\end{equation}
with $a_n^{(N)}, b^{(N)} \in \mathds{C}$.
This enables
 us to extract the resonance lying closest to the expansion point $s_0$. The pole position $s_\text{R}^{(N)}$ and corresponding residue $\mathcal{R}^{(N)}$ of the Pad\'e approximant are given by
\begin{align}
    s_\text{R}^{(N)}&=-\frac{1}{b^{(N)}}+s_0^{(N)}\,, \notag\\ \mathcal{R}^{(N)}&=\sum_{n=0}^N (-1)^{n}\frac{a_n^{(N)}}{b^{(N)}{}^{n+1}}\,.
\end{align}
For a more detailed introduction into the applications of Pad\'e theory, see Refs.~\cite{SanzCillero:2010mp, Masjuan:2013jha,Masjuan:2014psa}.

To determine the parameters $a_n^{(N)}$ and  $b^{(N)}$, we fit Pad\'e approximants to the scattering matrix $T_{11}$ and the form factor $(\FF)_1$. As both $T_{11}$ and $(\FF)_1$ contain the same pole, the parameter $b^{(N)}$ is the same, while the $a_n^{(N)}$ coefficients are allowed to be different
in the two fits. Note that in the present study the energy range of the form factor is limited by the
$\tau$ mass, such that the parameters
of the $\Kc$ resonance are fixed by the $T$-matrix only. Furthermore, the $\Kb$ resonance is located in close proximity to the $\ketap$ threshold, in such a way that the expansion of Eq.~\eqref{eq:pade_definition} needs to be modified.
To include the non-analyticities of the two closest relevant thresholds,
in Ref.~\cite{Masjuan:2013jha} it was proposed to perform the expansion
not in $s$, but in the conformal variable
\begin{equation}
\omega(s)=\frac{\sqrt{s-s^\text{th}_{1}}-\sqrt{s^\text{th}_{2}-s}}{\sqrt{s-s^\text{th}_{1}}+\sqrt{s^\text{th}_{2}-s}}\,,
\label{eq:omega_conf}
\end{equation}
with $s^\text{th}_{1}$ and $s^\text{th}_{2}$ denoting the lower and upper threshold, respectively. This transformation maps the first and adjacent unphysical Riemann sheet to a unit circle in $\omega$ without introducing any unphysical discontinuities, allowing for a better convergence of the Pad\'e series. As the main decay channel of the $\Kb$ is $\kpi$, we set $s^\text{th}_{1}=(\mk+\mpi)^2$ and $s^\text{th}_{2}=(\mk+\metap)^2$.

The systematic uncertainty originating from the Pad\'e approximant is estimated by
\begin{equation}
	\Delta_\text{sys}^{(N)} = \abs{\sqrt{s_\text{R}^{(N)}} - \sqrt{s_\text{R}^{(N-1)}}}\,.
	\label{eq:pade_uncertainty}
\end{equation}
The statistical uncertainty is obtained via a bootstrap analysis by varying the underlying amplitudes within their respective $1\sigma$ uncertainties. The resulting uncertainties are then added
in quadrature,
\begin{equation}
    \Delta^{(N)}_\text{total}=\sqrt{\Delta_\text{st}^2+{\Delta^{(N)}_\text{sys}}^2}\,.
\end{equation}
For any given value of $N$, the extracted pole position in general still depends on the expansion 
point $s_0^{(N)}$. Hence we first calculate the 
Pad\'e approximants for a wide range of $s_0^{(N)}$ values. For appropriate values of $s_0^{(N)}$ the extracted pole 
stabilizes. We therefore choose that value of $s_0^{(N)}$
for each $N$ that minimizes $\Delta_\text{sys}^{(N)}$. 
Furthermore, resonance couplings and branching ratios to $\kpi$ can be calculated via the residue $\mathcal{R}$, which can be expressed in terms of the $T$-matrix
\begin{equation}
    \lim_{s \rightarrow s_\text{p} } (s-s_\text{p}) T_{ij} = -\mathcal{R}_{ij}\,.
\end{equation}
For the normalization of the channel coupling $\tilde g^\text{R}_{j}$ of a resonance R to channel $j$ we choose the convention given in the resonance review of Ref.~\cite{Zyla:2020zbs}
\begin{equation}
    \tilde g^\text{R}_j=\mathcal{R}_{ij}/\sqrt{\mathcal{R}_{ii}}\,,
\end{equation}
corresponding to a partial width $\Gamma_{\text{R} \rightarrow i}$ defined by
\begin{equation}
\label{GRi}
    \Gamma_{\text{R} \rightarrow i}=\frac{|\tilde g^\text{R}_i|^2}{M_\text{R}} \rho_i(M_\text{R}^2)\,.
\end{equation}
Note that only for narrow, non-overlapping resonances, these partial widths sum up to the resonance width $\Gamma_\text{R}$ defined in Eq.~\eqref{eq:pole_defintion}.

\begin{figure}[tp]
	\includegraphics[width=\linewidth,trim = 18pt 19pt 0pt 0pt]{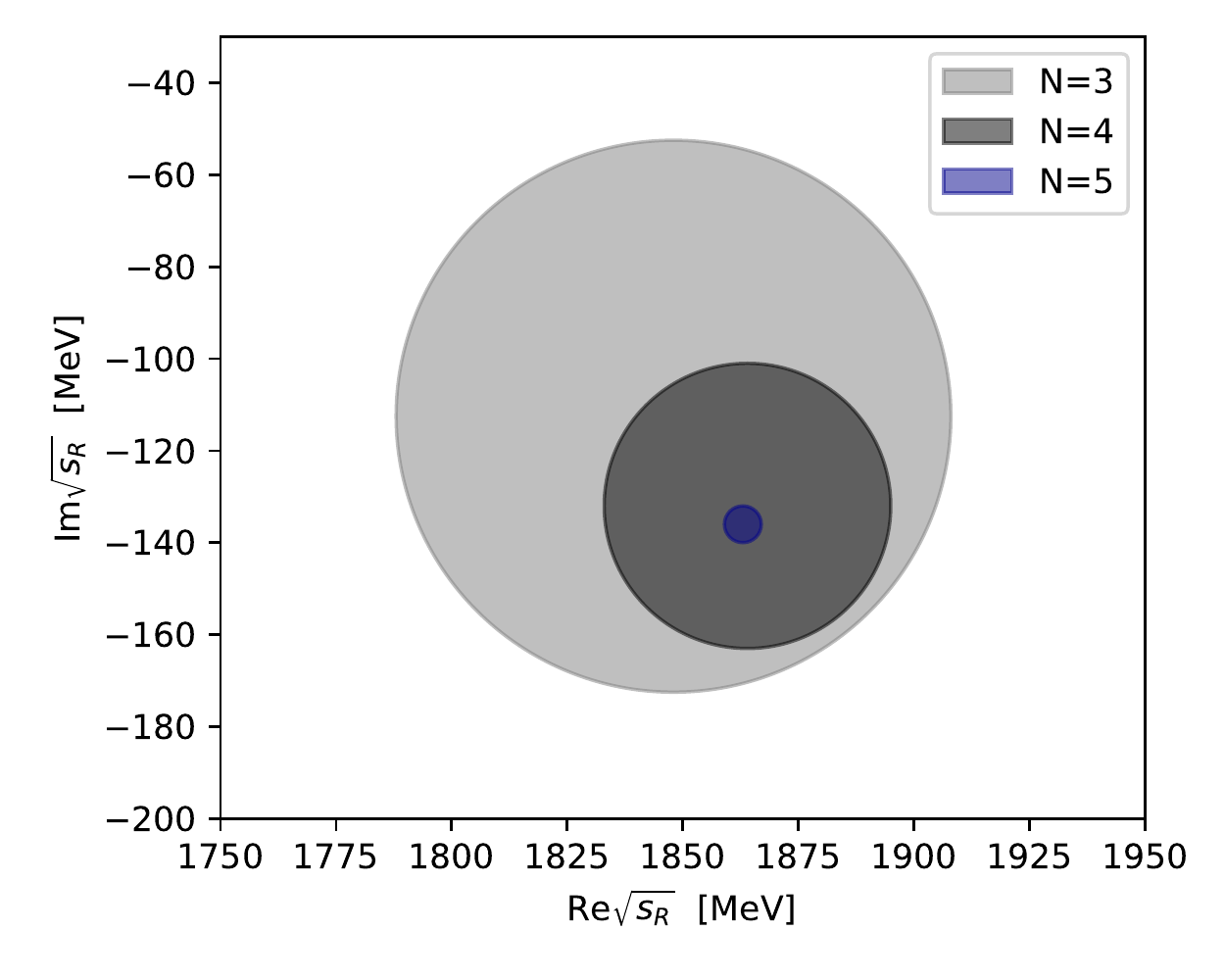}
	\centering
	\caption{Uncertainty regions $\Delta_\text{sys}^{(N)}$ as defined in Eq.~\eqref{eq:pade_uncertainty} for the optimal value of $s_0$ on the pole position of the $\Kc$.}
	\label{fig:pole_1950_2}
\end{figure}

\begin{figure}[tp]
	\includegraphics[width=\linewidth,trim = 18pt 19pt 0pt 0pt]{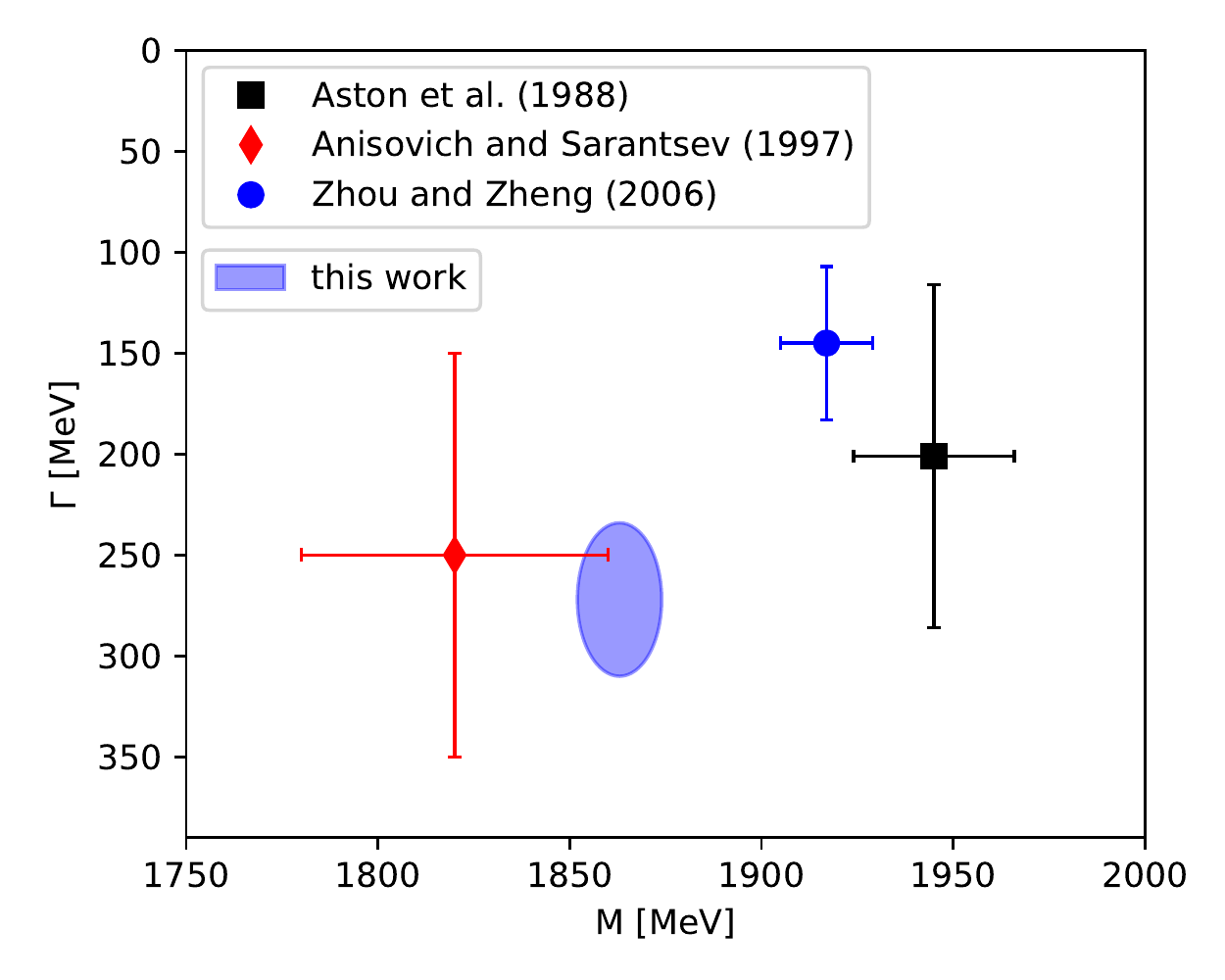}
	\centering
	\caption{Extracted pole position of the $\Kc$ in comparison to the works of Aston et al.~\cite{Aston:1987ir}, Anisovich and Sarantsev~\cite{Anisovich:1997qp}, and Zhou and Zheng~\cite{Zhou:2006wm}.}
	\label{fig:pole_1950}
\end{figure}

\begin{table}[tp]
\renewcommand{\arraystretch}{1.3}
	\begin{tabular}{l  r }
		\toprule
		$\sqrt{s_\text{R}^{(5)}}$~[$\text{MeV}$] & $1863(11)(4)-i~136(19)(4)$\\
		$\Delta_\text{sys}^{(5)}$~[$\text{MeV}$]  & $4$\\
		$\sqrt{s_0^{(5)}}$~[$\text{GeV}$] & $1.86$\\
		\midrule
		$\text{mod}\left(\tilde g^{\Kc}_{\kpi}\right)$~[$\text{GeV}$] & $4.32(35)(8)$\\
		$\text{arg}\left(\tilde g^{\Kc}_{\kpi}\right)$ & $-0.20(3)(1)$\\
		$\Gamma_{\Kc \rightarrow \kpi}$~[$\text{MeV}$] & $184(19)(4)$\\
		$\Gamma_{\Kc \rightarrow \kpi}$/$\Gamma_\text{tot}$ & $0.70(7)(2)$\\
		\bottomrule
	\end{tabular}
	\centering
	\caption{Results of Pad\'e extractions at ${N=5}$ for the $\Kc$ including statistical (first bracket) and systematic uncertainties (second bracket).}
	\label{tab:pole_1950}
\renewcommand{\arraystretch}{1.0}
\end{table}

For the $\Kc$, Fig.~\ref{fig:pole_1950_2} illustrates a clear convergence of the pole position for increasing orders $N$ of the Pad\'e series, which is truncated at ${N=5}$. 
The extracted pole corresponds to a mass of $1863(11)(4)\MeV$ and a decay width of $272(38)(8)\MeV$, as shown in Table~\ref{tab:pole_1950}. According to the PDG, the $\Kc$ still needs confirmation and is quoted with a mass of $1945(22)\MeV$ and a width of $201(90)\MeV$~\cite{Zyla:2020zbs}. However, this average is based only on the analysis of Ref.~\cite{Aston:1987ir}, where the data are fit using a simple BW distribution with an energy-dependent width and a linear background term in a limited energy range including the $\Kc$.
A comparison to the other results listed by the PDG is shown in Fig.~\ref{fig:pole_1950}. Both Refs.~\cite{Anisovich:1997qp,Zhou:2006wm}
are based on the data from Ref.~\cite{Aston:1987ir}. Reference~\cite{Anisovich:1997qp} uses a $K$-matrix formalism including $\kpi$, $\ketap$, and a generalized multimeson channel. The authors find, similarly to our analysis, a state with comparable width but a significantly smaller mass in comparison to Ref.~\cite{Aston:1987ir}. Using a BW ansatz with constant width, improved by incorporating some of the  left- and circular-cut contributions via a dispersion integral, Ref.~\cite{Zhou:2006wm} finds a $\Kc$ mass and width similar to Ref.~\cite{Aston:1987ir}, which should probably be expected given that the left-hand cut contribution should not influence the $1.9\GeV$ mass region significantly. 

We find a partial width to $\kpi$ of $184(19)(4)\MeV$, resulting in a branching fraction of $0.70(7)(2)$. This is slightly larger than the results obtained in Ref.~\cite{Aston:1987ir} with $0.52(8)(12)$ and the estimate in Ref.~\cite{Zhou:2006wm} quoting a branching fraction of $0.6$, however, still compatible within uncertainties. As the statistical uncertainties become most significant at higher energies, they dominate the total uncertainty.

\begin{figure}[tp]%
	\includegraphics[width=\linewidth,trim = 18pt 19pt 0pt 0pt]{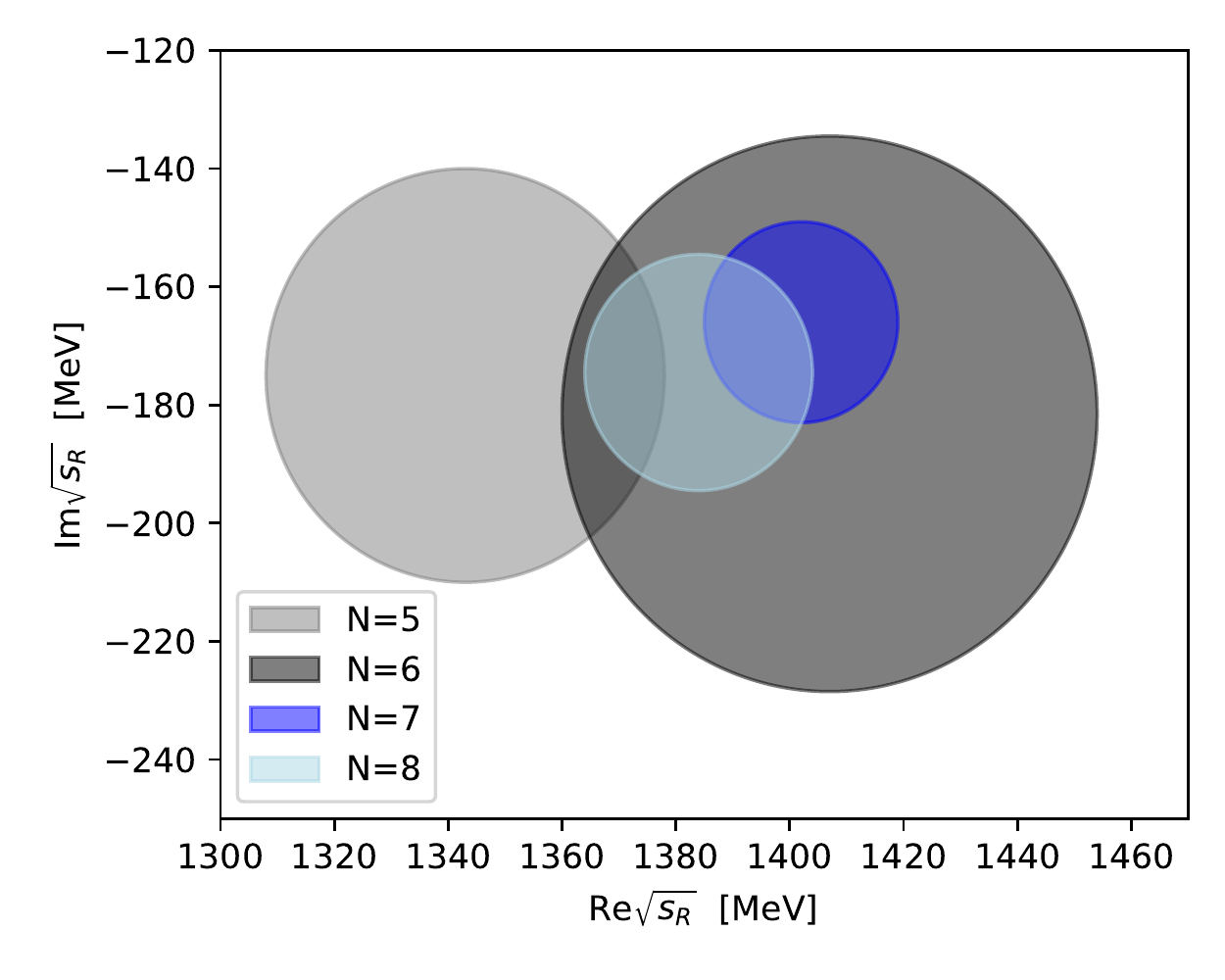}
	\centering
	\caption{Uncertainty regions $\Delta_\text{sys}^{(N)}$ as defined in Eq.~\eqref{eq:pade_uncertainty} for the optimal value of $s_0$ on the pole position of the $\Kb$.}
	\label{fig:pole_1430_2}
\end{figure}

\begin{figure}[tp]%
	\includegraphics[width=\linewidth,trim = 18pt 19pt 0pt 0pt]{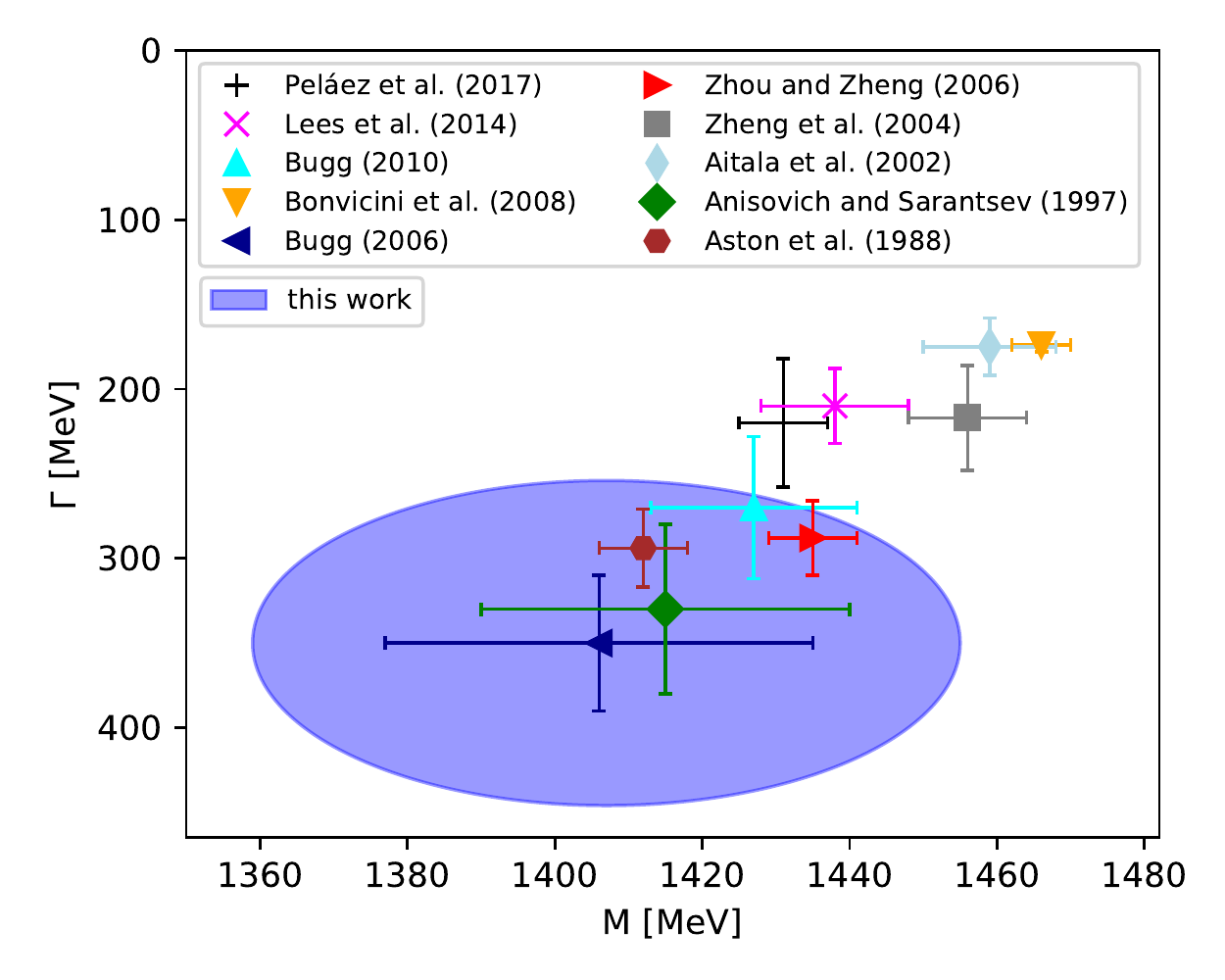}
	\centering
	\caption{Extracted pole position of the $\Kb$ in comparison to Pel\'aez et al.~\cite{Pelaez:2016klv}, Lees et al.~\cite{Lees:2014iua}, Bugg~(2010)~\cite{Bugg:2009uk}, Bonvicini et al.~\cite{Bonvicini:2008jw}, Bugg~(2006)~\cite{Bugg:2005xx}, Zhou  and Zheng~\cite{Zhou:2006wm}, Zeng et al.~\cite{Zheng:2003rw}, Aitala et al.~\cite{Aitala:2002kr}, Anisovich and Serantsev~\cite{Anisovich:1997qp}, and Aston et al.~\cite{Aston:1987ir}.}
	\label{fig:pole_1430}
\end{figure}

In the case of the $\Kb$ resonance, the Pad\'e fits prove less stable than for the $\Kc$. 
According to Montessus' theorem, to ensure convergence of the Pad\'e series, the amplitude $F(s)$ needs to be known in a compact subset of the disc $B_\delta(s_0)$.
Therefore, the Pad\'e approximants  should, in principle, also be fixed off the real axis.
However, for this procedure we observe large statistical uncertainties on the coefficients of the polynomial as well as the pole parameters, which are likely induced by the nearby  branch-point singularity of the $\ketap$ threshold. Alternatively fixing the Pad\'e approximants solely on the real axis dramatically improves the stability of the fit. Although the fit then has less information on the gradient in the complex plane, we still find a good agreement of this fit to the amplitude even off the real axis  on the physical sheet. Furthermore, this Pad\'e series reproduces $T$-matrix and form factor on the real axis, which is the only part fixed by experimental data, particularly well. We still observe an unusual fluctuation of the extracted pole position, especially for small orders $N$ of the Pad\'e approximants, as seen in Fig.~\ref{fig:pole_1430_2}. However, for ${N\geq6}$ the theoretical uncertainty begins to stabilize, with $\Delta_\text{sys}^{(N)}$ being consistent with the value of ${N=6}$. Since the ${N=7}$ and ${N=8}$ ellipses do not fully overlap and the uncertainty of ${N=8}$ is even larger than that of ${N=7}$, we use parameters and uncertainties derived from ${N=6}$ as our final outcome; see Table~\ref{tab:pole_1430}.
This results in a mass of $1408(4)(47)\MeV$ and a decay width of $360(14)(96)\MeV$. The comparison to other values listed by the PDG is shown in Fig.~\ref{fig:pole_1430}. 
 
\begin{table}[tp]
\renewcommand{\arraystretch}{1.3}
	\begin{tabular}{l r }
		\toprule
		$\sqrt{s_\text{R}^{(6)}}$~[$\text{MeV}$] & $1408(4)(47)-i ~ 180(7)(47)$\\
		$\Delta_\text{sys}^{(6)}$~[$\text{MeV}$]  &  $47$\\
		$\sqrt{s_0^{(6)}}$~[$\text{GeV}$] & $1.36$\\
		\midrule
		$\text{mod}\left(\tilde g^{\Kb}_{\kpi}\right)$~[$\text{GeV}$] & $4.96(14)(78)$\\
		$\text{arg}\left(\tilde g^{\Kb}_{\kpi}\right)$ & $0.06(1)(4)$\\
		$\Gamma_{\Kb \rightarrow \kpi}$~[$\text{MeV}$] & $304(8)(42)$\\
		$\Gamma_{\Kb \rightarrow \kpi}$/$\Gamma_\text{tot}$ & $0.87(2)(11)$\\
		\bottomrule
	\end{tabular}
	\centering
	\caption{Results of Pad\'e extractions at ${N=6}$ for the $\Kb$ including statistical (first bracket) and systematic uncertainties (second bracket).}
	\label{tab:pole_1430}
\renewcommand{\arraystretch}{1.0}
\end{table}

Although we use the elastic formalism of Ref.~\cite{Pelaez:2016tgi} as input, we obtain a somewhat larger decay width for the $\Kb$ than the one extracted in Ref.~\cite{Pelaez:2016klv}, although the difference is not significant.
It might be related to the fact that our fits of the phase shifts start to deviate from those of Ref.~\cite{Pelaez:2016tgi} in the energy range of the $\Kb$, see Fig.~\ref{fig:Fit_Aston}. 
As illustrated in Fig.~\ref{fig:pole_1430_2}, the Pad\'e analysis is quite stable with respect to the width of the $\Kb$, which gives some confidence in the rather large value of $\Gamma_{\Kb}$ extracted here.
For the partial width to $\kpi$ we find a value of $304(8)(42)\MeV$, corresponding to a branching fraction of $0.87(2)(11)$. Within uncertainties this is consistent with the value obtained in Ref.~\cite{Aston:1987ir}, $0.93(4)(9)$.

Finally, the coupling of a scalar resonance R to the $\bar s\gamma^\mu u$ current is again determined in a model-independent way in terms of its residue $C_\text{R}^{us}$, which can be extracted from the scalar form factor near the pole $s_\text{R}$ according to
\begin{equation}
\label{Cus}
    \bar f_0(s)=\sqrt{\frac{2}{3}}\frac{\tilde g^\text{R}_{\pi K} C_\text{R}^{us}}{s_\text{R}-s},
\end{equation}
where the coefficient has been chosen to ensure that $C_\text{R}^{us}$ matches the conventions of Refs.~\cite{Maltman:1999jn,ElBennich:2009da}.
The results for $C_{\Kb}^{us}$ for our Fits 1--4 are given in Table~\ref{tab:residue_1430}, compared to literature values from Refs.~\cite{Maltman:1999jn,ElBennich:2009da,Barate:1999hj}. We stress that $C_\text{R}^{us}$ is the unambiguous observable that describes the resonance properties, which only for narrow resonances corresponds to a physical branching fraction. However, to facilitate the comparison of different conventions, it is useful to formally define branching fractions by the narrow-width relation, even for resonances as broad as the $\sigma$~\cite{Moussallam:2011zg,Hoferichter:2011wk}. In the case of the $\bar s\gamma^\mu u$ current
the decay width for $\tau\to \text{R} \nu_\tau$ for an $S$-wave resonance R  in the narrow-width approximation reads
\begin{align}
\label{tauK0BR}
    \Gamma(\tau \to \text{R} \nu_\tau) 
    & = \frac{ 6 \pi^2 c_\Gamma \Delta_{\kpi}^2}{M_\text{R}^4} \Big(1-\frac{M_\text{R}^2}{m_\tau^2}\Big)^2\big|C_\text{R}^{us}\big|^2,
\end{align}
where the factor $\text{BR}(\text{R}\to\pi K)$ for the branching fraction should be added if R is only reconstructed in the $\pi K$ channel. This form can be extracted from Eq.~\eqref{eq:BelleDecaywidth} by inserting Eq.~\eqref{Cus}, identifying the square of the propagator with ${\pi/(M_\text{R}\Gamma_\text{R})\delta(s-M_\text{R}^2)}$, where $\Gamma_\text{R}$
is connected to $s_\text{R}$ via Eq.~\eqref{eq:pole_defintion} and the limit ${\Gamma_\text{R}\to 0}$ is formally
assumed, and finally expressing ${|\tilde g^\text{R}_{\pi K}|^2}$ by the resonance width defined in Eq.~\eqref{GRi} (see Refs.~\cite{Hoferichter:2012pm,Kubis:2015sga,Hoferichter:2017ftn} for more details, to identify the $\rho$ contribution in the pseudoscalar decays ${P\to\pi\pi\gamma}$ and crossed reactions). In addition, Eq.~\eqref{tauK0BR} includes a factor $3$ to account for all $K\pi$ decay channels of the $\Kb$. 

\begin{table}[t]
\renewcommand{\arraystretch}{1.3}
	\begin{tabular}{l r r}
		\toprule
		& $\big|C_{\Kb}^{us}\big|$~[$\text{GeV}$] & $\text{BR}(\tau\to \Kb \nu_\tau)$\\\midrule
		Fit~1 & $0.23(4)$ & $0.31(11)\times 10^{-4}$\\
		Fit~2 & $0.37(8)$ & $0.78(38)\times 10^{-4}$\\
		Fit~3 & $0.11(2)$ & $0.07(3)\times 10^{-4}$\\
		Fit~4 & $0.31(5)$ & $0.55(20)\times 10^{-4}$\\
		\midrule
\multicolumn{3}{l}{Other theoretical determinations:} \\
		Ref.~\cite{Maltman:1999jn} & $0.37$ & $0.79\times 10^{-4}$ \\
		Ref.~\cite{ElBennich:2009da} & $0.28$ & $0.45\times 10^{-4}$\\
		\midrule
\multicolumn{3}{l}{Experiment:} \\
		Ref.~\cite{Barate:1999hj} & $<0.93$ & $<5\times 10^{-4}$\\
		\bottomrule
	\end{tabular}
	\centering
	\caption{Results for the $\Kb$ residue and the corresponding ${\tau\to \Kb \nu_\tau}$ branching fraction, in comparison to the literature. The limit is given at $95\%$ confidence level.}
	\label{tab:residue_1430}
\renewcommand{\arraystretch}{1.0}
\end{table}

Table~\ref{tab:residue_1430} shows that the residues extracted from the fits to ${\tau\to K_S\pi\nu_\tau}$ scatter around the literature values from other theoretical investigations.
We deduce from our analysis an upper limit for the ${\tau\to \Kb \nu_\tau}$ branching
fraction, $\text{BR}(\tau\to \Kb \nu_\tau)<1.6 \times 10^{-4}$ (at $95\%$ confidence level), that improves
the literature value~\cite{Zyla:2020zbs,Barate:1999hj} by a factor 3. 
The difference of our findings to those of Refs.~\cite{Maltman:1999jn,ElBennich:2009da} can be traced back to their more rigid parameterization of the scalar form factor, which in both cases is determined once the input $T$-matrix is specified. In our case, when the coupling to the $\Kc$ or a linear term in the source are admitted, the fit to the $\tau$ spectrum implies a larger scalar form factor than in the most constrained Fit~3, which translates into a larger residue. 
A more reliable extraction of $|C_{\Kb}^{us}|$ from ${\tau\to K_S\pi\nu_\tau}$ would thus require more precise data, ideally on the forward--backward asymmetry to allow for a better separation of the $S$- and $P$-wave components. Such data actually exist for ${B\to K\pi\pi}$ decays~\cite{ElBennich:2009da}, but at the expense of additional hadronic uncertainties due to the presence of the spectator pion.

\section{Conclusions}
\label{sec:conclusions}

In this paper we presented a parameterization of the scalar $\kpi$ form factor that extends into the inelastic region, assuming that the main inelastic effects proceed via the $\Kb$ and $\Kc$ resonances. Technically, this is achieved by combining an Omn\`es description in terms of the $\kpi$ phase shift, valid in the elastic region, with a potential ansatz that incorporates (bare) resonance poles. 
The formalism, here employed for two channels, ensures that the result respects all constraints from analyticity and
unitarity. In particular, Watson's theorem in its domain of validity is fulfilled by construction.
We then collected the required input to determine 
the $S$-wave $\kpi$ $T$-matrix from state-of-the-art analyses of $\kpi$ scattering, leading to a parameterization
of the scalar form factor valid up to $2.3 \GeV$.

As a first application, we considered the $\tau\to K_S\pi\nu_\tau$ spectrum, especially its part above $1\GeV$ that is dominated by inelastic contributions both in the $S$- and $P$-wave. Employing a RChT description of the vector form factor, we found that, in combination with the information on the $\Kb$ incorporated in the input $S$-wave $T$-matrix, the spectrum allows one to determine the $\Ke$ mass at the level of $30\MeV$. An improved separation of these overlapping resonances, as well as distinguishing among the fit variants shown in Figs.~\ref{fig:Fit_Decay}--\ref{fig:Fit_VFF}, requires additional information on the form factor phase motions, which could be obtained via a future measurement of the forward--backward asymmetry, e.g., at Belle II, see Fig.~\ref{fig:AsymFB}. Estimating the $\Ke$ tensor coupling via large-$N_c$ arguments, the resulting vector form factor also allows us to derive a refined estimate of the $CP$ asymmetry generated by a tensor operator, see 
Eq.~\eqref{CPasym}.

Finally, we considered the extraction of the $\Kb$ and $\Kc$ resonance properties from the scalar $T$-matrix and form factor via Pad\'e approximants, leading to the constraints on the pole positions shown in Figs.~\ref{fig:pole_1950} and \ref{fig:pole_1430}. For the $\Kc$ we find an improved precision compared to previous analyses, with an uncertainty dominated by the fit to $\kpi$ scattering input. For the $\Kb$ the uncertainty is dominated by the systematics of the Pad\'e expansion, likely related to the proximity of the $\eta' K$ threshold. We also provided results for the pole residues and the corresponding branching fractions, both for $\pi K$ scattering and the coupling to 
the $\bar s \gamma^\mu u$ current.  

In conclusion, our representation for the scalar $\kpi$ form factor proves adequate well beyond the elastic region, and can be constrained phenomenologically up to energies where 
analyses of $\kpi$ scattering are available. This covers the entire kinematic range probed in the $\tau$ decay, includes the $\Kc$ resonance region, and should thus allow for a meaningful description of the $\kpi$ form factor in future analyses of semileptonic $D$- and $B$-meson decays. In particular, combined analyses with ${\tau\to K_S\pi\nu_\tau}$ data along the lines presented here should keep the number of free parameters to a minimum.

\begin{acknowledgements}
We thank D.~Epifanov and S.~Eidelman for communication on Ref.~\cite{Epifanov:2007rf}
and D.~van Dyk, J.~R.~Pel\'aez, A.~Rodas, S.~Ropertz, and J.~Ruiz de Elvira for useful discussions.
Financial support by the SNSF (Project No.\ PCEFP2\_181117),
the DFG through the funds provided to the Sino--German Collaborative
Research Center TRR110 ``Symmetries and the Emergence of Structure in QCD'' (DFG Project-ID 196253076 -- TRR 110),
the Bonn--Cologne Graduate School of Physics and Astronomy (BCGS),
and the European Union's Horizon 2020 research and innovation programme under grant agreement No.\ 824093
is gratefully acknowledged.
\end{acknowledgements}

\bibliographystyle{utphysmod.bst}
\balance
\bibliography{refs}

\providecommand{\href}[2]{#2}\begingroup\raggedright\begin{thebibliography}{100}

\bibitem{Bernard:1990kw}
V.~Bernard, N.~Kaiser, and U.-G. Mei{\ss}ner, Nucl. Phys. B {\bfseries 357},
  129 (1991).

\bibitem{Bijnens:2004bu}
J.~Bijnens, P.~Dhonte, and P.~Talavera, JHEP {\bfseries 05}, 036 (2004),
  [\href{https://arxiv.org/abs/hep-ph/0404150}{{arXiv:hep-ph/0404150}}].

\bibitem{Buettiker:2003pp}
P.~B{\"u}ttiker, S.~Descotes-Genon, and B.~Moussallam, Eur. Phys. J. C
  {\bfseries 33}, 409 (2004),
  [\href{https://arxiv.org/abs/hep-ph/0310283}{{arXiv:hep-ph/0310283}}].

\bibitem{DescotesGenon:2006uk}
S.~Descotes-Genon and B.~Moussallam, Eur. Phys. J. C {\bfseries 48}, 553
  (2006),
  [\href{https://arxiv.org/abs/hep-ph/0607133}{{arXiv:hep-ph/0607133}}].

\bibitem{Pelaez:2016klv}
J.~R. Pel\'aez, A.~Rodas, and J.~Ruiz~de Elvira, Eur. Phys. J. C {\bfseries
  77}, 91 (2017),
  [\href{https://arxiv.org/abs/1612.07966}{{arXiv:1612.07966~[hep-ph]}}].

\bibitem{Pelaez:2020uiw}
J.~R. Pel\'aez and A.~Rodas, Phys. Rev. Lett. {\bfseries 124}, 172001 (2020),
  [\href{https://arxiv.org/abs/2001.08153}{{arXiv:2001.08153~[hep-ph]}}].

\bibitem{Pelaez:2020gnd}
J.~R. Pel\'aez and A.~Rodas,
  \href{https://arxiv.org/abs/2010.11222}{{arXiv:2010.11222~[hep-ph]}}.

\bibitem{Pelaez:2021dak}
J.~R. Pel\'aez, A.~Rodas, and J.~Ruiz~de Elvira,
  \href{https://arxiv.org/abs/2101.06506}{{arXiv:2101.06506~[hep-ph]}}.

\bibitem{Pelaez:2018qny}
J.~R. Pel\'aez and A.~Rodas, Eur. Phys. J. C {\bfseries 78}, 897 (2018),
  [\href{https://arxiv.org/abs/1807.04543}{{arXiv:1807.04543~[hep-ph]}}].

\bibitem{Dax:2020dzg}
M.~Dax, D.~Stamen, and B.~Kubis, Eur. Phys. J. C {\bfseries 81}, 221 (2021),
  [\href{https://arxiv.org/abs/2012.04655}{{arXiv:2012.04655~[hep-ph]}}].

\bibitem{Colangelo:2015kha}
G.~Colangelo, E.~Passemar, and P.~Stoffer, Eur. Phys. J. C {\bfseries 75}, 172
  (2015),
  [\href{https://arxiv.org/abs/1501.05627}{{arXiv:1501.05627~[hep-ph]}}].

\bibitem{Niecknig:2015ija}
F.~Niecknig and B.~Kubis, JHEP {\bfseries 10}, 142 (2015),
[\href{https://arxiv.org/abs/1509.03188}{{arXiv:1509.03188~[hep-ph]}}].

\bibitem{Niecknig:2017ylb}
F.~Niecknig and B.~Kubis, Phys. Lett. {\bfseries B780}, 471 (2018),
[\href{https://arxiv.org/abs/1708.00446}{{arXiv:1708.00446~[hep-ph]}}].

\bibitem{Ditsche:2012fv}
C.~Ditsche, M.~Hoferichter, B.~Kubis, and U.-G. Mei{\ss}ner, JHEP {\bfseries
  06}, 043 (2012),
  [\href{https://arxiv.org/abs/1203.4758}{{arXiv:1203.4758~[hep-ph]}}].

\bibitem{Hoferichter:2015hva}
M.~Hoferichter, J.~Ruiz~de Elvira, B.~Kubis, and U.-G. Mei{\ss}ner, Phys. Rept.
  {\bfseries 625}, 1 (2016),
  [\href{https://arxiv.org/abs/1510.06039}{{arXiv:1510.06039~[hep-ph]}}].

\bibitem{Donoghue:1990xh}
J.~F. Donoghue, J.~Gasser, and H.~Leutwyler, Nucl. Phys. B {\bfseries 343}, 341
  (1990).

\bibitem{Hoferichter:2012wf}
M.~Hoferichter, C.~Ditsche, B.~Kubis, and U.-G. Mei{\ss}ner, JHEP {\bfseries
  06}, 063 (2012),
  [\href{https://arxiv.org/abs/1204.6251}{{arXiv:1204.6251~[hep-ph]}}].

\bibitem{Watson:1954uc}
K.~M. Watson, Phys. Rev. {\bfseries 95}, 228 (1954).

\bibitem{Bernard:2006gy}
V.~Bernard, M.~Oertel, E.~Passemar, and J.~Stern, Phys. Lett. B {\bfseries
  638}, 480 (2006),
  [\href{https://arxiv.org/abs/hep-ph/0603202}{{arXiv:hep-ph/0603202}}].

\bibitem{Bernard:2009zm}
V.~Bernard, M.~Oertel, E.~Passemar, and J.~Stern, Phys. Rev. D {\bfseries 80},
  034034 (2009),
  [\href{https://arxiv.org/abs/0903.1654}{{arXiv:0903.1654~[hep-ph]}}].

\bibitem{Abouzaid:2009ry}
E.~Abouzaid {\em et~al.} [KTeV], Phys. Rev. D {\bfseries 81}, 052001 (2010),
  [\href{https://arxiv.org/abs/0912.1291}{{arXiv:0912.1291~[hep-ex]}}].

\bibitem{Moussallam:2007qc}
B.~Moussallam, Eur. Phys. J. C {\bfseries 53}, 401 (2008),
  [\href{https://arxiv.org/abs/0710.0548}{{arXiv:0710.0548~[hep-ph]}}].

\bibitem{Boito:2008fq}
D.~R. Boito, R.~Escribano, and M.~Jamin, Eur. Phys. J. C {\bfseries 59}, 821
  (2009), [\href{https://arxiv.org/abs/0807.4883}{{arXiv:0807.4883~[hep-ph]}}].

\bibitem{Boito:2010me}
D.~R. Boito, R.~Escribano, and M.~Jamin, JHEP {\bfseries 09}, 031 (2010),
  [\href{https://arxiv.org/abs/1007.1858}{{arXiv:1007.1858~[hep-ph]}}].

\bibitem{Bernard:2011ae}
V.~Bernard, D.~R. Boito, and E.~Passemar, Nucl. Phys. B Proc. Suppl. {\bfseries
  218}, 140 (2011),
  [\href{https://arxiv.org/abs/1103.4855}{{arXiv:1103.4855~[hep-ph]}}].

\bibitem{Antonelli:2013usa}
M.~Antonelli, V.~Cirigliano, A.~Lusiani, and E.~Passemar, JHEP {\bfseries 10},
  070 (2013),
  [\href{https://arxiv.org/abs/1304.8134}{{arXiv:1304.8134~[hep-ph]}}].

\bibitem{Omnes:1958hv}
R.~Omnès, Nuovo Cim. {\bfseries 8}, 316 (1958).

\bibitem{Ecker:1988te}
G.~Ecker, J.~Gasser, A.~Pich, and E.~de~Rafael, Nucl. Phys. B {\bfseries 321},
  311 (1989).

\bibitem{Oller:1998zr}
J.~A. Oller and E.~Oset, Phys. Rev. D {\bfseries 60}, 074023 (1999),
  [\href{https://arxiv.org/abs/hep-ph/9809337}{{arXiv:hep-ph/9809337}}].

\bibitem{Jamin:2000wn}
M.~Jamin, J.~A. Oller, and A.~Pich, Nucl. Phys. B {\bfseries 587}, 331 (2000),
  [\href{https://arxiv.org/abs/hep-ph/0006045}{{arXiv:hep-ph/0006045}}].

\bibitem{Noel:2020vpo}
F.~No\"el, L.~von Detten, and C.~Hanhart, Bull. Lebedev Phys. Inst. {\bfseries
  47}, 330 (2020).

\bibitem{Callan:1966hu}
C.~G. Callan and S.~B. Treiman, Phys. Rev. Lett. {\bfseries 16}, 153 (1966).

\bibitem{Dashen:1969bh}
R.~F. Dashen and M.~Weinstein, Phys. Rev. Lett. {\bfseries 22}, 1337 (1969).

\bibitem{Gasser:1984ux}
J.~Gasser and H.~Leutwyler, Nucl. Phys. B {\bfseries 250}, 517 (1985).

\bibitem{Bijnens:2007xa}
J.~Bijnens and K.~Ghorbani,
  \href{https://arxiv.org/abs/0711.0148}{{arXiv:0711.0148~[hep-ph]}}.

\bibitem{Kastner:2008ch}
A.~Kastner and H.~Neufeld, Eur. Phys. J. C {\bfseries 57}, 541 (2008),
  [\href{https://arxiv.org/abs/0805.2222}{{arXiv:0805.2222~[hep-ph]}}].

\bibitem{Cirigliano:2017tqn}
V.~Cirigliano, A.~Crivellin, and M.~Hoferichter, Phys. Rev. Lett. {\bfseries
  120}, 141803 (2018),
  [\href{https://arxiv.org/abs/1712.06595}{{arXiv:1712.06595~[hep-ph]}}].

\bibitem{Ablikim:2015mjo}
M.~Ablikim {\em et~al.} [BESIII], Phys. Rev. D {\bfseries 94}, 032001 (2016),
  [\href{https://arxiv.org/abs/1512.08627}{{arXiv:1512.08627~[hep-ex]}}].

\bibitem{Oller:2004xm}
J.~A. Oller, Phys. Rev. D {\bfseries 71}, 054030 (2005),
  [\href{https://arxiv.org/abs/hep-ph/0411105}{{arXiv:hep-ph/0411105}}].

\bibitem{ElBennich:2006yi}
B.~El-Bennich, A.~Furman, R.~Kami\'nski, L.~Le\'sniak, and B.~Loiseau, Phys.
  Rev. D {\bfseries 74}, 114009 (2006),
  [\href{https://arxiv.org/abs/hep-ph/0608205}{{arXiv:hep-ph/0608205}}].

\bibitem{ElBennich:2009da}
B.~El-Bennich, A.~Furman, R.~Kami\'nski, L.~Le\'sniak, B.~Loiseau, and
  B.~Moussallam, Phys. Rev. D {\bfseries 79}, 094005 (2009),
  [\href{https://arxiv.org/abs/0902.3645}{{arXiv:0902.3645~[hep-ph]}}],
  [Erratum: Phys. Rev. D {\bf 83}, 039903 (2011)].

\bibitem{Boito:2009qd}
D.~R. Boito and R.~Escribano, Phys. Rev. D {\bfseries 80}, 054007 (2009),
  [\href{https://arxiv.org/abs/0907.0189}{{arXiv:0907.0189~[hep-ph]}}].

\bibitem{Boito:2017jav}
D.~Boito, J.-P. Dedonder, B.~El-Bennich, R.~Escribano, R.~Kami\'nski,
  L.~Le\'sniak, and B.~Loiseau, Phys. Rev. D {\bfseries 96}, 113003 (2017),
  [\href{https://arxiv.org/abs/1709.09739}{{arXiv:1709.09739~[hep-ph]}}].

\bibitem{Mizuk:2009da}
R.~Mizuk {\em et~al.} [Belle], Phys. Rev. D {\bfseries 80}, 031104 (2009),
  [\href{https://arxiv.org/abs/0905.2869}{{arXiv:0905.2869~[hep-ex]}}].

\bibitem{Aaij:2014jqa}
R.~Aaij {\em et~al.} [LHCb], Phys. Rev. Lett. {\bfseries 112}, 222002 (2014),
  [\href{https://arxiv.org/abs/1404.1903}{{arXiv:1404.1903~[hep-ex]}}].

\bibitem{Ropertz:2018stk}
S.~Ropertz, C.~Hanhart, and B.~Kubis, Eur. Phys. J. C {\bfseries 78}, 1000
  (2018),
  [\href{https://arxiv.org/abs/1809.06867}{{arXiv:1809.06867~[hep-ph]}}].

\bibitem{Hanhart:2012wi}
C.~Hanhart, Phys. Lett. B {\bfseries 715}, 170 (2012),
  [\href{https://arxiv.org/abs/1203.6839}{{arXiv:1203.6839~[hep-ph]}}].

\bibitem{Daub:2015xja}
J.~T. Daub, C.~Hanhart, and B.~Kubis, JHEP {\bfseries 02}, 009 (2016),
  [\href{https://arxiv.org/abs/1508.06841}{{arXiv:1508.06841~[hep-ph]}}].

\bibitem{Nakano:1982bc}
K.~Nakano, Phys. Rev. C {\bfseries 26}, 1123 (1982).

\bibitem{Anisovich:2002ij}
V.~V. Anisovich and A.~V. Sarantsev, Eur. Phys. J. A {\bfseries 16}, 229
  (2003),
  [\href{https://arxiv.org/abs/hep-ph/0204328}{{arXiv:hep-ph/0204328}}].

\bibitem{Klempt:2007cp}
E.~Klempt and A.~Zaitsev, Phys. Rept. {\bfseries 454}, 1 (2007),
  [\href{https://arxiv.org/abs/0708.4016}{{arXiv:0708.4016~[hep-ph]}}].

\bibitem{Battaglieri:2014gca}
M.~Battaglieri {\em et~al.}, Acta Phys. Polon. B {\bfseries 46}, 257 (2015),
  [\href{https://arxiv.org/abs/1412.6393}{{arXiv:1412.6393~[hep-ph]}}].

\bibitem{Aitchison:1972ay}
I.~J.~R. Aitchison, Nucl. Phys. A {\bfseries 189}, 417 (1972).

\bibitem{Zyla:2020zbs}
P.~A. Zyla {\em et~al.} [Particle Data Group], PTEP {\bfseries 2020}, 083C01
  (2020).

\bibitem{Khuri:1960zz}
N.~N. Khuri and S.~B. Treiman, Phys. Rev. {\bfseries 119}, 1115 (1960).

\bibitem{Baru:2020ywb}
V.~Baru, E.~Epelbaum, A.~A. Filin, C.~Hanhart, R.~V. Mizuk, A.~V. Nefediev, and
  S.~Ropertz, Phys. Rev. D {\bfseries 103}, 034016 (2021),
  [\href{https://arxiv.org/abs/2012.05034}{{arXiv:2012.05034~[hep-ph]}}].

\bibitem{Amaryan:2020xhw}
M.~Amaryan {\em et~al.} [KLF],
  \href{https://arxiv.org/abs/2008.08215}{{arXiv:2008.08215~[nucl-ex]}}.

\bibitem{Bakker:1970wg}
A.~M. Bakker {\em et~al.}, Nucl. Phys. B {\bfseries 24}, 211 (1970).

\bibitem{Cho:1970fb}
Y.~Cho {\em et~al.}, Phys. Lett. B {\bfseries 32}, 409 (1970).

\bibitem{Estabrooks:1977xe}
P.~Estabrooks, R.~K. Carnegie, A.~D. Martin, W.~M. Dunwoodie, T.~A. Lasinski,
  and D.~W. G.~S. Leith, Nucl. Phys. B {\bfseries 133}, 490 (1978).

\bibitem{Jongejans:1973pn}
B.~Jongejans, R.~A. van Meurs, A.~G. Tenner, H.~Voorthuis, P.~M. Heinen, W.~J.
  Metzger, H.~G. J.~M. Tiecke, and R.~T. Van~de Walle, Nucl. Phys. B {\bfseries
  67}, 381 (1973).

\bibitem{Linglin:1973ci}
D.~Linglin {\em et~al.}, Nucl. Phys. B {\bfseries 57}, 64 (1973).

\bibitem{Aston:1987ir}
D.~Aston {\em et~al.}, Nucl. Phys. B {\bfseries 296}, 493 (1988).

\bibitem{Pelaez:2016tgi}
J.~R. Pel\'aez and A.~Rodas, Phys. Rev. D {\bfseries 93}, 074025 (2016),
  [\href{https://arxiv.org/abs/1602.08404}{{arXiv:1602.08404~[hep-ph]}}].

\bibitem{Epifanov:2007rf}
D.~Epifanov {\em et~al.} [Belle], Phys. Lett. B {\bfseries 654}, 65 (2007),
  [\href{https://arxiv.org/abs/0706.2231}{{arXiv:0706.2231~[hep-ex]}}].

\bibitem{Tishchenko:2012ie}
V.~Tishchenko {\em et~al.} [MuLan], Phys. Rev. D {\bfseries 87}, 052003 (2013),
  [\href{https://arxiv.org/abs/1211.0960}{{arXiv:1211.0960~[hep-ex]}}].

\bibitem{Marciano:1985pd}
W.~J. Marciano and A.~Sirlin, Phys. Rev. Lett. {\bfseries 56}, 22 (1986).

\bibitem{Marciano:1988vm}
W.~J. Marciano and A.~Sirlin, Phys. Rev. Lett. {\bfseries 61}, 1815 (1988).

\bibitem{Braaten:1990ef}
E.~Braaten and C.-S. Li, Phys. Rev. D {\bfseries 42}, 3888 (1990).

\bibitem{Antonelli:2009ws}
M.~Antonelli {\em et~al.}, Phys. Rept. {\bfseries 494}, 197 (2010),
  [\href{https://arxiv.org/abs/0907.5386}{{arXiv:0907.5386~[hep-ph]}}].

\bibitem{Gasser:1984gg}
J.~Gasser and H.~Leutwyler, Nucl. Phys. B {\bfseries 250}, 465 (1985).

\bibitem{Jamin:2001zq}
M.~Jamin, J.~A. Oller, and A.~Pich, Nucl. Phys. B {\bfseries 622}, 279 (2002),
  [\href{https://arxiv.org/abs/hep-ph/0110193}{{arXiv:hep-ph/0110193}}].

\bibitem{Aoki:2019cca}
S.~Aoki {\em et~al.} [Flavour Lattice Averaging Group], Eur. Phys. J. C
  {\bfseries 80}, 113 (2020),
  [\href{https://arxiv.org/abs/1902.08191}{{arXiv:1902.08191~[hep-lat]}}].

\bibitem{Jamin:2008qg}
M.~Jamin, A.~Pich, and J.~Portol{\'e}s, Phys. Lett. B {\bfseries 664}, 78
  (2008), [\href{https://arxiv.org/abs/0803.1786}{{arXiv:0803.1786~[hep-ph]}}].

\bibitem{Lepage:1979zb}
G.~P. Lepage and S.~J. Brodsky, Phys. Lett. B {\bfseries 87}, 359 (1979).

\bibitem{Lepage:1980fj}
G.~P. Lepage and S.~J. Brodsky, Phys. Rev. D {\bfseries 22}, 2157 (1980).

\bibitem{Beldjoudi:1994hi}
L.~Beldjoudi and T.~N. Truong, Phys. Lett. B {\bfseries 351}, 357 (1995),
  [\href{https://arxiv.org/abs/hep-ph/9411423}{{arXiv:hep-ph/9411423}}].

\bibitem{Kou:2018nap}
W.~Altmannshofer {\em et~al.} [Belle-II], PTEP {\bfseries 2019}, 123C01 (2019),
  [\href{https://arxiv.org/abs/1808.10567}{{arXiv:1808.10567~[hep-ex]}}],
  [Erratum: PTEP {\bf 2020}, 029201 (2020)].

\bibitem{Ryu:2014vpc}
S.~Ryu {\em et~al.} [Belle], Phys. Rev. D {\bfseries 89}, 072009 (2014),
  [\href{https://arxiv.org/abs/1402.5213}{{arXiv:1402.5213~[hep-ex]}}].

\bibitem{BABAR:2011aa}
J.~P. Lees {\em et~al.} [BaBar], Phys. Rev. D {\bfseries 85}, 031102 (2012),
  [\href{https://arxiv.org/abs/1109.1527}{{arXiv:1109.1527~[hep-ex]}}],
  [Erratum: Phys. Rev. D {\bf 85}, 099904 (2012)].

\bibitem{Grossman:2011zk}
Y.~Grossman and Y.~Nir, JHEP {\bfseries 04}, 002 (2012),
  [\href{https://arxiv.org/abs/1110.3790}{{arXiv:1110.3790~[hep-ph]}}].

\bibitem{Devi:2013gya}
H.~Z. Devi, L.~Dhargyal, and N.~Sinha, Phys. Rev. D {\bfseries 90}, 013016
  (2014), [\href{https://arxiv.org/abs/1308.4383}{{arXiv:1308.4383~[hep-ph]}}].

\bibitem{Dhargyal:2016jgo}
L.~Dhargyal, Springer Proc. Phys. {\bfseries 203}, 329 (2018),
  [\href{https://arxiv.org/abs/1610.06293}{{arXiv:1610.06293~[hep-ph]}}].

\bibitem{Rendon:2019awg}
J.~Rend\'on, P.~Roig, and G.~Toledo~S\'anchez, Phys. Rev. D {\bfseries 99},
  093005 (2019),
  [\href{https://arxiv.org/abs/1902.08143}{{arXiv:1902.08143~[hep-ph]}}].

\bibitem{Dighe:2019odu}
A.~Dighe, S.~Ghosh, G.~Kumar, and T.~S. Roy,
  \href{https://arxiv.org/abs/1902.09561}{{arXiv:1902.09561~[hep-ph]}}.

\bibitem{Chen:2019vbr}
F.-Z. Chen, X.-Q. Li, Y.-D. Yang, and X.~Zhang, Phys. Rev. D {\bfseries 100},
  113006 (2019),
  [\href{https://arxiv.org/abs/1909.05543}{{arXiv:1909.05543~[hep-ph]}}].

\bibitem{Cata:2008zc}
O.~Cat\`a and V.~Mateu, Phys. Rev. D {\bfseries 77}, 116009 (2008),
  [\href{https://arxiv.org/abs/0801.4374}{{arXiv:0801.4374~[hep-ph]}}].

\bibitem{Baum:2011rm}
I.~Baum, V.~Lubicz, G.~Martinelli, L.~Orifici, and S.~Simula, Phys. Rev. D
  {\bfseries 84}, 074503 (2011),
  [\href{https://arxiv.org/abs/1108.1021}{{arXiv:1108.1021~[hep-lat]}}].

\bibitem{Hoferichter:2018zwu}
M.~Hoferichter, B.~Kubis, J.~Ruiz~de Elvira, and P.~Stoffer, Phys. Rev. Lett.
  {\bfseries 122}, 122001 (2019),
  [\href{https://arxiv.org/abs/1811.11181}{{arXiv:1811.11181~[hep-ph]}}],
  [Erratum: Phys. Rev. Lett. {\bf 124}, 199901 (2020)].

\bibitem{SanzCillero:2010mp}
J.~J. Sanz-Cillero,
  \href{https://arxiv.org/abs/1002.3512}{{arXiv:1002.3512~[hep-ph]}}.

\bibitem{Masjuan:2013jha}
P.~Masjuan and J.~J. Sanz-Cillero, Eur. Phys. J. C {\bfseries 73}, 2594 (2013),
  [\href{https://arxiv.org/abs/1306.6308}{{arXiv:1306.6308~[hep-ph]}}].

\bibitem{Masjuan:2014psa}
P.~Masjuan, J.~Ruiz~de Elvira, and J.~J. Sanz-Cillero, Phys. Rev. D {\bfseries
  90}, 097901 (2014),
  [\href{https://arxiv.org/abs/1410.2397}{{arXiv:1410.2397~[hep-ph]}}].

\bibitem{Anisovich:1997qp}
A.~V. Anisovich and A.~V. Sarantsev, Phys. Lett. B {\bfseries 413}, 137 (1997),
  [\href{https://arxiv.org/abs/hep-ph/9705401}{{arXiv:hep-ph/9705401}}].

\bibitem{Zhou:2006wm}
Z.~Y. Zhou and H.~Q. Zheng, Nucl. Phys. A {\bfseries 775}, 212 (2006),
  [\href{https://arxiv.org/abs/hep-ph/0603062}{{arXiv:hep-ph/0603062}}].

\bibitem{Lees:2014iua}
J.~P. Lees {\em et~al.} [BaBar], Phys. Rev. D {\bfseries 89}, 112004 (2014),
  [\href{https://arxiv.org/abs/1403.7051}{{arXiv:1403.7051~[hep-ex]}}].

\bibitem{Bugg:2009uk}
D.~V. Bugg, Phys. Rev. D {\bfseries 81}, 014002 (2010),
  [\href{https://arxiv.org/abs/0906.3992}{{arXiv:0906.3992~[hep-ph]}}].

\bibitem{Bonvicini:2008jw}
G.~Bonvicini {\em et~al.} [CLEO], Phys. Rev. D {\bfseries 78}, 052001 (2008),
  [\href{https://arxiv.org/abs/0802.4214}{{arXiv:0802.4214~[hep-ex]}}].

\bibitem{Bugg:2005xx}
D.~V. Bugg, Phys. Lett. B {\bfseries 632}, 471 (2006),
  [\href{https://arxiv.org/abs/hep-ex/0510019}{{arXiv:hep-ex/0510019}}].

\bibitem{Zheng:2003rw}
H.~Q. Zheng, Z.~Y. Zhou, G.~Y. Qin, Z.~Xiao, J.~J. Wang, and N.~Wu, Nucl. Phys.
  A {\bfseries 733}, 235 (2004),
  [\href{https://arxiv.org/abs/hep-ph/0310293}{{arXiv:hep-ph/0310293}}].

\bibitem{Aitala:2002kr}
E.~M. Aitala {\em et~al.} [E791], Phys. Rev. Lett. {\bfseries 89}, 121801
  (2002),
  [\href{https://arxiv.org/abs/hep-ex/0204018}{{arXiv:hep-ex/0204018}}].

\bibitem{Maltman:1999jn}
K.~Maltman, Phys. Lett. B {\bfseries 462}, 14 (1999),
  [\href{https://arxiv.org/abs/hep-ph/9906267}{{arXiv:hep-ph/9906267}}].

\bibitem{Barate:1999hj}
R.~Barate {\em et~al.} [ALEPH], Eur. Phys. J. C {\bfseries 11}, 599 (1999),
  [\href{https://arxiv.org/abs/hep-ex/9903015}{{arXiv:hep-ex/9903015}}].

\bibitem{Moussallam:2011zg}
B.~Moussallam, Eur. Phys. J. C {\bfseries 71}, 1814 (2011),
  [\href{https://arxiv.org/abs/1110.6074}{{arXiv:1110.6074~[hep-ph]}}].

\bibitem{Hoferichter:2011wk}
M.~Hoferichter, D.~R. Phillips, and C.~Schat, Eur. Phys. J. C {\bfseries 71},
  1743 (2011),
  [\href{https://arxiv.org/abs/1106.4147}{{arXiv:1106.4147~[hep-ph]}}].

\bibitem{Hoferichter:2012pm}
M.~Hoferichter, B.~Kubis, and D.~Sakkas, Phys. Rev. D {\bfseries 86}, 116009
  (2012), [\href{https://arxiv.org/abs/1210.6793}{{arXiv:1210.6793~[hep-ph]}}].

\bibitem{Kubis:2015sga}
B.~Kubis and J.~Plenter, Eur. Phys. J. C {\bfseries 75}, 283 (2015),
  [\href{https://arxiv.org/abs/1504.02588}{{arXiv:1504.02588~[hep-ph]}}].

\bibitem{Hoferichter:2017ftn}
M.~Hoferichter, B.~Kubis, and M.~Zanke, Phys. Rev. D {\bfseries 96}, 114016
  (2017),
  [\href{https://arxiv.org/abs/1710.00824}{{arXiv:1710.00824~[hep-ph]}}].

\end{thebibliography}\endgroup

\end{document}